\documentclass[oneside,english,letterpaper]{aa}
\usepackage[OT1]{fontenc}
\usepackage[latin1]{inputenc}
\usepackage{amsmath}
\usepackage{graphicx}
\usepackage{amssymb}
\usepackage[authoryear]{natbib}

\makeatletter

\providecommand{\tabularnewline}{\\}

\usepackage{graphicx}
\usepackage{amssymb}

\newcommand{\dd}{\mathrm d}
\newcommand{\trig}{\mathrm{trig}}
\newcommand{\ee}{\mathrm e}
\newcommand{\ii}{\mathrm i}
\newcommand{\boldfwm}{\boldsymbol{\mathcal{F}}}
\newcommand{\fwm}{{\mathcal{F}}}

\renewcommand{\cite}{\citep}

\bibpunct{(}{)}{;}{a}{}{,}

\usepackage{babel}
\makeatother
\begin{document}

\title{The cosmic shear three-point functions}

\author{K.~Benabed \inst{1} \and R.~Scoccimarro \inst{2}}

\institute{Institut d'Astrophysique de Paris, 98bis bd Arago, 75014 Paris, France
\and Center for Cosmology and Particule Physics, Department of Physics,
New York University, New York, NY 10003, USA. }

\abstract{We investigate the three-point functions of the weak lensing cosmic
shear, using both analytic methods and numerical results from N-body
simulations. The analytic model, an isolated dark matter halo with
a power-law profile chosen to fit the effective index at the scale
probed, can be used to understand the basic properties of the eight
three-point functions observed in simulations. We use this model to
construct a single three-point function estimator that {}``optimally''
combines the eight three-point functions. This new estimator is an
alternative to $M_{ap}$ statistics and provides up to a factor of
two improvement in signal to noise compared to previously used combinations
of cosmic shear three-point functions. }

\maketitle

\section{Introduction}

The quality of recent and upcoming galaxy weak lensing surveys is
rapidly improving and allows increasingly high signal to noise determination
of second and third order cosmic shear correlation functions, which
contain very interesting cosmological information. For example, the
two point function constrains a combination of $\Omega_{m}$ and $\sigma_{8}$
\cite{1997AA...322....1B} with some sensitivity to the shape of the
primordial power spectrum \cite{2002AA...396....1S} and the equation
of state of the dark energy component \cite{2003astro.ph..6033B}.
The three-point function, which we investigate in this paper, is particularly
important for breaking degeneracies on two-point statistics, giving
a strong constraint on $\Omega_{m}$ \cite{1997AA...322....1B} with
a small dependency on the dark energy component \cite{2001PhRvD..64h3501B}.
Studies of three-point statistics have recently received a great deal
of attention from the theoretical side \cite{2003AA...397..405B,2003AA...397..809S,2003ApJ...583L..49T,2003astro.ph..4034T,2002MNRAS.337..875T,2003ApJ...584..559Z,2005A&A...431....9S},
inspired in part by the detection of an averaged cosmic shear three-point
function \cite{2002AA...389L..28B}.

Weak gravitational lensing detection is based on studying the alignment
of background galaxies due to the lensing effect of the intervening
mass \cite{Bartelmann:1999yn}, which provides a measure of the average
shear $\gamma$. From this estimator, one could in principle obtain
the projected mass $\kappa$, however this requires inverting a non-local
relation (see Eq. (\ref{eq:gammadef}) below), which is very sensitive
to the boundary conditions and thus difficult in galaxy surveys with
complicated geometries and masks due to bright stars, etc.

One possible alternative is to build statistics of the projected mass
$\kappa$ without reconstruction of $\kappa$ itself, by finding a
local operation on the shear that will give some filtered version
of the projected mass. A well-known procedure of this type, the aperture
mass statistic or $M_{ap}$ \cite{2002AA...396....1S}, corresponds
to averaging the two-point function of $\kappa$ with a compensated
(zero integrated volume) filter. In terms of the shear, it corresponds
to convolving the shear two-point functions with a compact support
filter. This statistic is not optimal in terms of signal to noise,
and although it can be extended to the three-point correlation function
\cite{2003AA...397..809S}, it leads to a determination of the projected
skewness with a rather poor signal to noise \cite{2003astro.ph..7393J,2003astro.ph..2031P}.
This can be solved partly by suitably designing the window involved
in the $M_{ap}$ statistic, but at the price of more complicated equations
\cite{2003astro.ph..7393J}.

For these reasons one would like to use statistics of the shear field
to constrain cosmological models. The difficulty with this approach
is that the shear is a spin-2 field, and thus the geometrical properties
of its correlation functions are much more complicated than for the
scalar field $\kappa$, particularly when one goes beyond two-point
statistics. For a general triangle configuration, there are eight
non-vanishing three-point functions of the cosmic shear field \cite{2002AA...396....1S,2003ApJ...583L..49T}.

In this work we study the main features of the cosmic shear three-point
functions using numerical simulations (see \cite{2003ApJ...583L..49T}
for previous work), and compare these results to a simple analytic
model (following the ideas in \cite{2003ApJ...584..559Z}) to extract
the main features for the purpose of defining an optimal linear combination
of the eight three-point functions into a single object, which is
easier to deal with. This generalizes the previous study in \cite{2003AA...397..405B},
where a particular linear combination was selected based on the pattern
of the three-point functions under some conditions, tested empirically
with numerical simulations. Here we pay particular attention to avoiding
cancellations that can occur when doing such linear combinations,
with the purpose of maximizing signal to noise.

This paper is organized as follows. In section \ref{sec:deux} we
discuss the basics of weak lensing cosmic shear, and present the geometrical
properties of the shear two-point functions \cite{1992ApJ...388..272K}
and then extend the results to the three-point functions \cite{2002AA...396....1S,2003ApJ...584..559Z}.
In section \ref{sec:Analyticalpredictions} we present a simple analytic
model based on a power-law profile, which allows us to extract the
basic properties of the shear three-point functions. In section \ref{sec:cinq}
we present the results of measurements in numerical simulations and
compare to the analytic model. In particular, we show that in the
range of scales between 1 and 3 arc minutes they agreement with an
isothermal sphere is very good. In section \ref{sec:estim} we use
these results to construct an estimator that combines the eight three-point
functions into a single object, and compare the signal to noise of
this new statistic to the estimator previously used in the literature
to detect a cosmic shear three-point function \cite{2002AA...389L..28B,2003AA...397..405B}.

\section{Weak lensing and three point function\label{sec:deux}}

\subsection{The Convergence and Shear fields}

To first order in the density perturbations, the weak lensing effect
is determined completely by its convergence field $\kappa$ which
is a simple projection of the matter density contrast along the line
of sight \cite{Bartelmann:1999yn,1997AA...322....1B}\begin{equation}
\kappa(\boldsymbol{\theta})=\int\frac{\dd z}{H(z)}\, w(z)\,\delta(\boldsymbol{\theta},z),\label{eq:kappadef}\end{equation}
 where the weight $w$ is the lensing efficiency function which reads
\begin{equation}
w(z)=\frac{3}{2}\Omega_{0}\:\frac{\mathrm{D}(z)\mathrm{D}(z,z_{s})}{\mathrm{D}(z_{s})}\:(1+z),\label{eq:wlensdef}\end{equation}
 in the case of a infinitely thin source plane at redshift $z_{s}$.
Here $\mathrm{D}(z_{1},z_{2})$ is the angular distances between redshifts
$z_{1}$ and $z_{2}$ and $\Omega_{m}$ the present dark matter density
in units of the critical density. We will focus our analysis on the
three-point function of the lensing shear $\boldsymbol{\gamma}$.
Ignoring higher-order corrections perturbation theory, the shear field
is simply related to the convergence field $\kappa$ by the non-local
equations\begin{equation}
\begin{array}{rcl}
\gamma_{1} & = & \Delta^{-1}\left(\partial_{1}^{2}-\partial_{2}^{2}\right)\kappa,\\
\gamma_{2} & = & \Delta^{-1}2\,\partial_{1}\partial_{2}\,\kappa,\end{array}\label{eq:gammadef}\end{equation}
 where $\Delta^{-1}$ denotes the inverse Laplacian operator. With
these equations, the information contained in the weak lensing effect
is simple to understand. The convergence field is simply a projection
along the line of sight of the matter density contrast. Thus its second
and third moments are related, up to projection effects, to the two-
and three-point functions of the density contrast. The shear and convergence
field, being equivalent through Eq. (\ref{eq:gammadef}), contain
the same information.

The shear $\boldsymbol{\gamma}$ transforms as a spin-2 object. Equation
(\ref{eq:gammadef}) is the analogous to that defining the $E$ and
$B$ components of the polarization in terms of the $Q$ and $U$
Stockes parameters. The convergence field plays here the role of the
$E$ polarization, and there is no $B$ component to the weak lensing
effect to this order. Such contribution can be produced by deviations
of the Born approximation, and have been shown to be more than two
orders of magnitude smaller than the dominant scalar mode \cite{2002ApJ...574...19C,2000ApJ...530..547J}.
In addition, in observations $B$ modes can arise due to systematic
effects. Recent detections of the cosmic shear have measured different
levels of $B$ modes (for a discussion see e.g. \cite{2003astro.ph..5089V}),
whose dominant source appears to be errors in the PSF correction.
Fortunately, recent improvements in this matter \cite{2003astro.ph..6097H,2004astro.ph.12234J,2005astro.ph..2243J}
result in a very low level of $B$ modes. Another potential systematics
is the possible intrinsic alignment of neighboring galaxies \cite{2002MNRAS.333..501B,2001MNRAS.320L...7C,2001ApJ...559..552C,2002ApJ...568...20C},
which can be ameliorated by using galaxies in different redshift bins
\cite{2003MNRAS.339..711H,2002AA...396..411K}, or in the specific
case of shear three-point functions, by taking advantage of its geometrical
properties \cite{2003astro.ph..5240S}.

\subsection{The shear two-point function\label{sub:shear2}}

The naive computation of the two point function of the shear field,
$\left\langle \gamma_{i}\gamma_{j}\right\rangle $, vanishes by symmetry
reasons. This is because the pseudo-vector $\boldsymbol{\gamma}$
transforms under rotation as a spin-2 object, which means that,\begin{equation}
\gamma=\gamma_{1}+\mathrm{i}\gamma_{2},\; R_{\phi}\left(\vec{\gamma}\right)=\left(\gamma_{1}+\mathrm{i}\gamma_{2}\right)\mathrm{e}^{\mathrm{i}2\phi},\end{equation}
 where $R_{\phi}$ represents a rotation by angle $\phi$. It is possible,
however, to build combinations of $\gamma_{i}\gamma_{j}$ that do
not average to zero. For example, the combination $\langle\gamma_{A}\gamma_{B}^{*}\rangle_{|AB|=\ell}$
is invariant under rotations, and thus non-zero upon averaging. More
generally, the two quantities \begin{eqnarray}
\xi_{+}(\theta) & = & \frac{1}{2}\left\langle \gamma_{A}\gamma_{B}^{*}+\gamma_{A}^{*}\gamma_{B}\right\rangle _{\left|AB\right|=\theta}\label{eq:xiplusdef}\\
\xi_{-}(\theta) & = & \frac{1}{2}\left\langle \gamma_{A}\gamma_{B}+\gamma_{A}^{*}\gamma_{B}^{*}\right\rangle _{\left|AB\right|=\theta},\label{eq:ximoinsdef}\end{eqnarray}
 are non-zero.

This property is easy to understand in terms of a \emph{tangential}
and \emph{cross} decomposition of the two-point shear correlation
function. If we call $\varphi$ the angle of the $AB$ vector relative
to the $x$ axis, we can define, for our two point a \emph{tangential}
($+$) and a \emph{cross} ($\times$) shear by\begin{equation}
\gamma_{+}+\mathrm{i}\gamma_{\times}=\left(\gamma_{1}+\mathrm{i}\gamma_{2}\right)\mathrm{e}^{-\mathrm{i}2\varphi},\end{equation}
 where the cross shear is oriented toward the direction $\pi/4+\varphi$,
to account for the transformation properties of the shear. In this
language the geometrical properties of the component of the shear
are easier to understand. The tangential component $\gamma_{+}$ is
of positive parity, whereas $\gamma_{\times}$ of negative parity.
We call parity transformation any reflexion, for example, the reflexion
on the $x$ axis that transform $y$ into $-y$. The reflexion about
the center of the segment $AB$ that exchange $A$ and $B$ is another
example that we will use in the following. Under parity transformation,
a $\gamma_{+}$ will transform into itself, whereas a $\gamma_{\times}$
will acquire a minus sign.

For the two-point function, a reflexion about the center of $AB$
is equal to a rotation by an angle $\pi$ due to the transformation
properties of the shear pseudo vector under rotations. Therefore,
for a parity negative two points function $D_{-}$, one would have,
for a parity transformation $P$\begin{eqnarray}
P(D_{-}) & = & -D_{-}\nonumber \\
 & = & R_{\pi}(D_{-})\nonumber \\
 & = & D_{-}\nonumber \\
 & = & 0.\label{eq:paritytwopoints}\end{eqnarray}
 Only the positive parity combinations of $\gamma_{+,\times}$ will
be non-zero. In other word, any non zero two-point function can be
decomposed as a linear combinations of \begin{eqnarray}
\xi_{++}(\theta) & = & \left\langle \gamma_{+}\gamma_{+}\right\rangle _{\theta}\\
\xi_{\times\times}(\theta) & = & \left\langle \gamma_{\times}\gamma_{\times}\right\rangle _{\theta}.\end{eqnarray}
 In particular the correlation functions $\xi_{\pm}$ defined above
read\begin{equation}
\xi_{\pm}(\theta)=\xi_{++}\pm\xi_{\times\times}.\end{equation}

\subsection{The shear three-point functions \label{sub:3pt:def}}

\begin{figure}
\begin{center}\includegraphics[%
  scale=0.4]{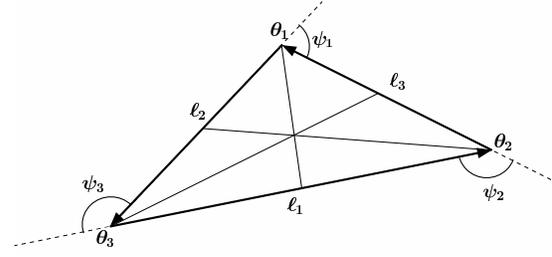}\end{center}

\caption{Definition and convention of triangle variables.\label{cap:triangledef}}
\end{figure}

Here we extend the approach in the previous section to the three-point
functions. In order to do this, we first need to define how we describe
triplets of points. For any triplet, we call the three vertices $\boldsymbol{\theta}_{i}$,
with $i$ taken modulo $3$, so that the triangles $\boldsymbol{\theta}_{0}\boldsymbol{\theta}_{1}\boldsymbol{\theta}_{2}$
is always oriented. We define the sides $\boldsymbol{\ell}_{i}$ and
the oriented angles at each vertex $\psi_{i}$ such that \begin{eqnarray}
\boldsymbol{\ell}_{i} & \equiv & \boldsymbol{\theta}_{i+1}-\boldsymbol{\theta}_{i+2}\\
\psi_{i} & \equiv & \left(\widehat{\boldsymbol{\ell}_{i+1},\boldsymbol{\ell}_{i+2}}\right).\end{eqnarray}
 Figure \ref{cap:triangledef} summarize these definitions.%

Because of translation invariance, the three-point functions depend
only on the relative positions of the vertices, so we shall use the
three lengths $\psi_{i}$ to specify our oriented triangles.

To define the shear three-point functions we decompose the shear into
cross and tangential components in some basis relative to the triangle
\cite{2003ApJ...584..559Z,2003AA...397..809S}. This is simpler than
trying to find all possible combinations of the three pseudo-vectors
$\boldsymbol{\gamma}$ that have non-zero three-point functions, but
of course is equivalent. Choosing a basis was {}``natural'' in the
case of the two-point function (given by the line joining the two
points), but less obvious for a triangle.

Since we are interested in the three-point functions, the knowledge
of the position of the points is irrelevant; only their relative positions
matters. Therefore, we will drop the vector notation for the vertices
$\boldsymbol{\theta}_{i}$. Thus, any configuration is completely
described by either the three points relative positions, the three
angles or the three side length. Of course, since we have decided
to only deal with oriented triangles, the length, and not the side
vectors are enough.

A simple solution is to pick some {}``special'' points of the triangle
and to project the shear along the vector linking this point and the
vertex we are interested in. Note that it is not, in fact, necessary
to choose such a special point. Any set of three orientations defined
by invariant properties of the configuration will do. In the following,
we call this choice of orientation a \emph{projection convention}.
Any change of projection convention can be described by a rotation
of each of the projection directions at each vertex of the triangle.
Of course, each of these rotations can in principle depend on the
shape of the triangle. We will call such change a \emph{rotation of
the projection convention} of angle $(\zeta_{1},\zeta_{2},\zeta_{3})$.

Once the projection convention is chosen, the shear field at the three
vertices of the triangle can be decomposed in a tangential and a cross
component, $\boldsymbol{\gamma}=(\gamma_{+},\gamma_{\times})$. As
in the two point function case, these two components have, respectively,
positive and negative parity. This decomposition leads to eight different
three-point functions $\gamma_{\mu\nu\rho}$ , $\mu$, $\nu$ and
$\rho$ being $+$ or $\times$ . Half of them, $\gamma_{+++}$ and
$\gamma_{+\times\times}$ plus permutations are parity positive, whereas
$\gamma_{\times\times\times}$ and $\gamma_{\times++}$ plus permutations
are of negative parity. Since the three-point functions also depend
upon the projection convention, we will sometimes denote them by $\gamma_{\mu\nu\rho}^{(\zeta_{1},\zeta_{2},\zeta_{3})}$,
$(\zeta_{1},\zeta_{2},\zeta_{3})$ being a rotation to the reference
frame.

We have seen that for the two-point function, parity properties reduce
the number of independent functions from four to only two. One can
wonder whether the parity properties will also reduce the number of
non zero three-point functions. The parity negative two-point functions
vanish because after a parity transformation, a simple rotation will
bring back the points to the initial position {[}see Eq. (\ref{eq:paritytwopoints}){]}.
This is not the case for three-point functions, except in the special
cases of equilateral triangles, and some isoceles configurations,
where a rotation can mimic a parity transformation \cite{2003AA...397..809S}.
This is why in the general case parity-negative three-point functions
are non-zero \cite{2003AA...397..809S,2003ApJ...583L..49T}. Formally
a parity transformation of a parity negative three-point function
for a configuration $\theta_{1}\theta_{2}\theta_{3}$ gives\begin{eqnarray}
P\left[\gamma_{\mu\nu\rho}^{(\zeta_{1},\zeta_{2},\zeta_{3})}(\ell_{1},\ell_{2},\ell_{3})\right] & = & -\gamma_{\mu\nu\rho}^{(\zeta_{1},\zeta_{2},\zeta_{3})}(\ell_{1},\ell_{2},\ell_{3})\nonumber \\
 & = & \gamma_{\mu\rho\nu}^{(\zeta_{1},\zeta_{3},\zeta_{2})}(\ell_{1},\ell_{3},\ell_{2}),\end{eqnarray}
 where we have assumed that the transformation exchange the last two
vertices. Thus, only if $\ell_{3}=\ell_{2}$, $\zeta_{2}(\ell_{1},\ell_{2},\ell_{3})=\zeta_{3}(\ell_{1},\ell_{2},\ell_{3})$
and $\rho=\nu$, parity forces the three-point function to be zero.
Note that for simple projection conventions, the property $\zeta_{2}(\ell_{1},\ell_{2},\ell_{3})=\zeta_{3}(\ell_{1},\ell_{2},\ell_{3})$
will usually be verified.

A free parameter that we overlooked in the case of two-point function
is the choice of projection convention. Indeed, we picked the easiest
projection where the angles from the reference frame were the same
for the two points. If we now allow these two angles to be different,
we break the property that the $\langle\gamma_{+}\gamma_{\times}\rangle$
are zero, since we cannot equate parity transformations and $\pi$
rotations, and we are then in similar situation to that of three-point
functions. The choice of projection that seemed most natural for the
two-point function has this extra property that its parity negative
component is zero.

Is it possible to find an analogous {}``natural'' projection convention
that reduces the number of non-zero three-point functions? This problem
has been partly investigated by \cite{2003AA...397..809S}. Here we
shall reproduce their results and extend them to answer the question
of the preferred projection convention.

In the two-point function cases, as we discussed above, a rotation
from the natural projection convention will induce non-zero parity
negative part. One can show that the $\xi_{\pm}$ will transform as\begin{equation}
\xi_{\pm}^{(\zeta_{1},\zeta_{2})}=e^{-i(\zeta_{1}\pm\zeta_{2})}\xi_{\pm}.\end{equation}
 In other words, the $\xi_{\pm}$ only acquire a phase, their amplitude
is invariant. One can build equivalent {}``rotational invariant''
the three-point functions as described in details in \cite{2003AA...397..809S},
where it is showed that any change of the projections axis of the
three points will leave the following four complex quantities invariants
up to a phase\begin{equation}
\Gamma_{0}=\left\langle \gamma_{1}\gamma_{2}\gamma_{3}\right\rangle ,\,\,\,\,\,\,\,\,\Gamma_{i}=\left\langle \gamma_{i}^{*}\prod_{j\neq i}\gamma_{j}\right\rangle ,\end{equation}
 with $i=1,2,3$, which can be written as, \begin{equation}
\Gamma_{\mu}=A_{\mu\nu}\gamma_{\nu}^{\oplus}-i\,\, B_{\mu\nu}\gamma_{\nu}^{\otimes},\end{equation}
 where $\mu,\nu=0,1,2,3$ and the matrices $A$ and $B$ have all
elements equal to $\pm1$, $B_{\mu\nu}\equiv1-2\delta_{\mu\nu}$,
$A_{\mu\nu}\equiv B_{\mu\nu}(1-2\delta_{\mu0})$. In addition, $\gamma^{\oplus}$
and $\gamma^{\otimes}$ denote, respectively, the positive- and negative-parity
three-point functions,\begin{equation}
\gamma^{\oplus}=\left(\begin{array}{c}
\gamma_{+++}\\
\gamma_{+\times\times}\\
\gamma_{\times+\times}\\
\gamma_{\times\times+}\end{array}\right),\,\,\,\,\,\,\,\,\,\,\gamma^{\otimes}=\left(\begin{array}{c}
\gamma_{\times\times\times}\\
\gamma_{\times++}\\
\gamma_{+\times+}\\
\gamma_{++\times}\end{array}\right).\end{equation}
 The transformation of the $\Gamma_{i}$ under the rotation of the
projection directions are given by\begin{eqnarray}
\Gamma'_{0} & = & \ee^{2\ii\left(\zeta_{1}+\zeta_{2}+\zeta_{3}\right)}\Gamma_{0}\label{eq:Gamma0Rot}\\
\Gamma'_{1} & = & \ee^{2\ii\left(-\zeta_{1}+\zeta_{2}+\zeta_{3}\right)}\Gamma_{1}\label{eq:Gamma1Rot}\\
\Gamma'_{2} & = & \ee^{2\ii\left(\zeta_{1}-\zeta_{2}+\zeta_{3}\right)}\Gamma_{2}\label{eq:Gamma2Rot}\\
\Gamma'_{3} & = & \ee^{2\ii\left(\zeta_{1}+\zeta_{2}-\zeta_{3}\right)}\Gamma_{3}.\label{eq:Gamma3Rot}\end{eqnarray}
 This property has some interesting consequences. Indeed, Eqs. (\ref{eq:Gamma0Rot}-\ref{eq:Gamma3Rot})
show that projection convention rotations will not change the modulus
of the $\Gamma_{i}$. At most, if a preferred projection exists (i.e.
if the eight three-point functions are redundant), they can be reduced
to four quantities. It might be possible that for a given triangle
shape one can find a preferred convention projection for which all
parity-negative three-point functions will be zero (one of course
can always find a rotation that yield \emph{three} vanishing three-point
functions). Let us note $(\zeta_{1},\zeta_{2},\zeta_{3})$ the rotation
that will transform a given projection convention into the projection
we are looking for. Eqs (\ref{eq:Gamma0Rot}-\ref{eq:Gamma3Rot})
define a system of four equations that our $\zeta_{i}$ have to verify.
This is possible if and only if $\Phi_{i}$, the phases of the $\Gamma_{i}$,
verify the relation\begin{equation}
\Phi_{0}=\Phi_{1}+\Phi_{2}+\Phi_{3}+k\pi\: k\epsilon\mathbb{Z}.\label{eq:simpleprojectionncondition}\end{equation}

For isosceles triangles, this condition can be shown to be true. Indeed,
the parity properties imply that for any isoceles configuration in
$\theta_{1}$, the following relations hold \begin{eqnarray}
\Gamma_{2} & = & \Gamma_{3}^{*}\\
\Gamma_{1} & = & \Gamma_{1}^{*}\\
\Gamma_{0} & = & \Gamma_{0}^{*},\end{eqnarray}
 that is, $\Phi_{0}$ and $\Phi_{1}$ are zero (modulo $\pi$), whereas
$\Phi_{2}=-\Phi_{1}$. The rotations that change a projection convention
to one with the zero parity negative three-point functions are of
the form $(0,-\Phi/4,\Phi/4)$, where $\Phi=\Phi_{2}=-\Phi_{3}$ is
the phase in the first projection convention. Note that this result
is of little use, since it just shows in a different language that
only four quantities are needed to describe the isosceles configurations,
which can be shown only from parity considerations. Moreover, the
phase $\Phi$ has to be evaluated in order to find the preferred convention
projection.

Neither parity properties, nor geometrical considerations allow us
to prove or disprove the conditions in Eq. (\ref{eq:simpleprojectionncondition})
in the general case. The answer to our question will have to come
from the computation of the three-point functions that we will exhibit
in the next section.

\section{Analytical predictions for the three point functions\label{sec:Analyticalpredictions}}

We now proceed to calculate the eight shear three-point functions.
This can be done in terms of the eight $\gamma_{\mu\nu\rho}$ or four
$\Gamma_{i}$ which are completely equivalent. In fact we will present
results in both representations, although most of the time we deal
with $\Gamma_{i}$.

To compute the three point functions of the shear, we will use a simple
model that captures most of the features of a more detailed calculation.
The reason for studying such a simplified model will become clear
in section \ref{sec:cinq} when we discuss how to {}``optimally''
combine the eight three-point functions into a single three-point
function.

\subsection{A single halo}

Weak lensing surveys provide their best constraints at scales small
enough (one to ten arc minutes) that are well into the non-linear
regime. For example, for measurements on background galaxies at redshift
around unity, the weak lensing efficiency is maximum at about $z=0.43$
for an $\Omega_{m}=0.3$, $\Omega_{\Lambda}=0.7$ cosmology, for which
one arc minute corresponds to a distance of 0.3 Mpc/h. For such scales,
contributions to lensing statistics come mainly from light deflection
by single dark matter halos, and it is a good approximation to compute
the light deflection ignoring coupling between different deflections
\cite{2002ApJ...574...19C,2001MNRAS.322..918V}. In the language of
the so-called halo model \cite{2002PhR...372....1C}, statistics are
dominated by the {}``one-halo'' contribution, and this has been
verified for the shear three-point functions by comparison with measurements
in numerical simulations \cite{2003ApJ...583L..49T,2003astro.ph..4034T}.
Remarkably, as shown in \cite{2003ApJ...584..559Z}, for scales of
about three arc minutes the full dependence of the shear three-point
functions on the triangle shape for the halo model agree very well
with a calculation based on a singular isothermal sphere up to an
overall amplitude. Given these results, we will only slightly go beyond
the singular isothermal model. We shall assume that the shear three-point
functions can be calculated by the contribution from one spherical
halo located at the maximum of the lensing efficiency window. The
halo profile will be taken as a general power-law

\begin{equation}
\rho(r)\propto\frac{1}{r^{n}},\label{eq:profile:simple}\end{equation}
 where $n=2$ corresponds to an isothermal sphere. In general, simulations
suggest dark matter halos have $n\sim1$ near the halo center and
$n\sim$3 at large distances \cite{1997ApJ...490..493N}, \begin{equation}
\rho\propto\frac{1}{r(r_{o}+r)^{2}},\label{eq:profile:gen}\end{equation}
 where the effective profile index is $n_{\textrm{eff}}=2$ when $r=r_{o},$typically
a tenth of the halo virial radius. Assuming a fixed $n$ (which can
be determined by the angular scale probed and the halo mass of the
dominant contribution to the three-point function) we loose cosmological
information, but on the other hand the calculation can be done analytically
and also the scaling properties of Eq. (\ref{eq:profile:simple})
allows us to ignore the halo profile normalization and the effects
of projection which enter as an overall constant.

Computing lensing by a spherical halo is very simple. From Eqs. (\ref{eq:kappadef}),
(\ref{eq:gammadef}) and (\ref{eq:profile:simple}) it follows that
the shear pattern behaves as\begin{equation}
{\boldsymbol{\gamma}}=\left[\begin{array}{c}
\cos(2\theta)\\
\sin(2\theta)\end{array}\right]\gamma(r),\end{equation}
 with\begin{equation}
\gamma(r)=r^{1-n}.\label{eq:shear:rn}\end{equation}
 The contribution of such a halo to the three-point functions is therefore\begin{eqnarray}
\gamma_{\mu\nu\rho}(\ell_{1},\ell_{2},\ell_{3}) & =\label{eq:gammamunurhodef}\\
 &  & \hspace{-2cm}\int\dd^{2}u\;\boldsymbol{P}_{\mu}\cdot\boldsymbol{\gamma}(\boldsymbol{\theta_{1}})\,\boldsymbol{P}_{\nu}\cdot\boldsymbol{\gamma}(\boldsymbol{\theta_{2}})\,\boldsymbol{P}_{\rho}\cdot\boldsymbol{\gamma}(\boldsymbol{\theta_{3}}),\nonumber \end{eqnarray}
 where the symbols $\boldsymbol{P}_{\mu}$ stand for the projectors
in the $+$ and $\times$ directions. The three points are taken at
the positions $\boldsymbol{\theta}_{i}$, such that $\boldsymbol{\ell}_{i}=\boldsymbol{\theta}_{i+1}-\boldsymbol{\theta}_{i+2}$,
and we use all our indices $i$ modulo 3. Note that we integrate out
the position of the center of the halo, denoted by $\boldsymbol{u}$.

Equations (\ref{eq:shear:rn}) and (\ref{eq:gammamunurhodef}) imply
that in our simple model, the following scaling relationship holds,
\begin{equation}
\gamma_{\mu\nu\rho}(\lambda\,\ell_{1},\lambda\,\ell_{2},\lambda\,\ell_{3})=\lambda^{5-3n}\,\gamma_{\mu\nu\rho}(\ell_{1},\ell_{2},\ell_{3}).\label{eq:scaling}\end{equation}

Since all choices of projections are equally valid, we can choose
here the projection operator to simplify the calculation as much as
possible. Here we will take the projection direction at each vertex
to be that of the opposite side. Note that this is equivalent, up
to a sign, to projecting along the lines joining each vertex of the
triangle to its orthocenter \cite{2003AA...397..809S}. For simplicity,
we shall refer to our projection as {}``orthocenter''. With this
choice, the projected shear reads\begin{equation}
\boldsymbol{P}_{\mu}\cdot\boldsymbol{\gamma}(\boldsymbol{\theta_{i}})=\frac{\trig_{\mu}(2\phi_{i})}{\left|\boldsymbol{\theta}_{i}-\boldsymbol{u}\right|^{n-1}},\label{eq:projectoppositedef}\end{equation}
 where $\trig_{+}\equiv\cos$ and $\trig_{\times}\equiv\sin$. The
angle $\phi_{i}$ is the angle defined by the line joining the vertex
and the center of the halo and the opposite side of the vertex\begin{equation}
\cos(\phi_{i})=\frac{\left(\boldsymbol{u}-\boldsymbol{\theta}_{i}\right)\cdot\boldsymbol{\ell}_{i}}{\left|\boldsymbol{u}-\boldsymbol{\theta}_{i}\right|\ell_{i}}.\end{equation}
 Given our choice of projection, the trigonometric functions can be
written in terms of the configuration $(\ell_{1},\ell_{2},\ell_{3})$,
and the relevant integrals computed in terms of hypergeometric functions
(see \cite{2003ApJ...584..559Z} for description of the integration
procedure). We refer the reader to Appendix A for details.

\subsection{Results}

\begin{figure}
\begin{tabular}{c}
 \includegraphics[
  bb=60bp 60bp 780bp 530bp,
  clip,
  width=1.0\columnwidth,
  keepaspectratio]{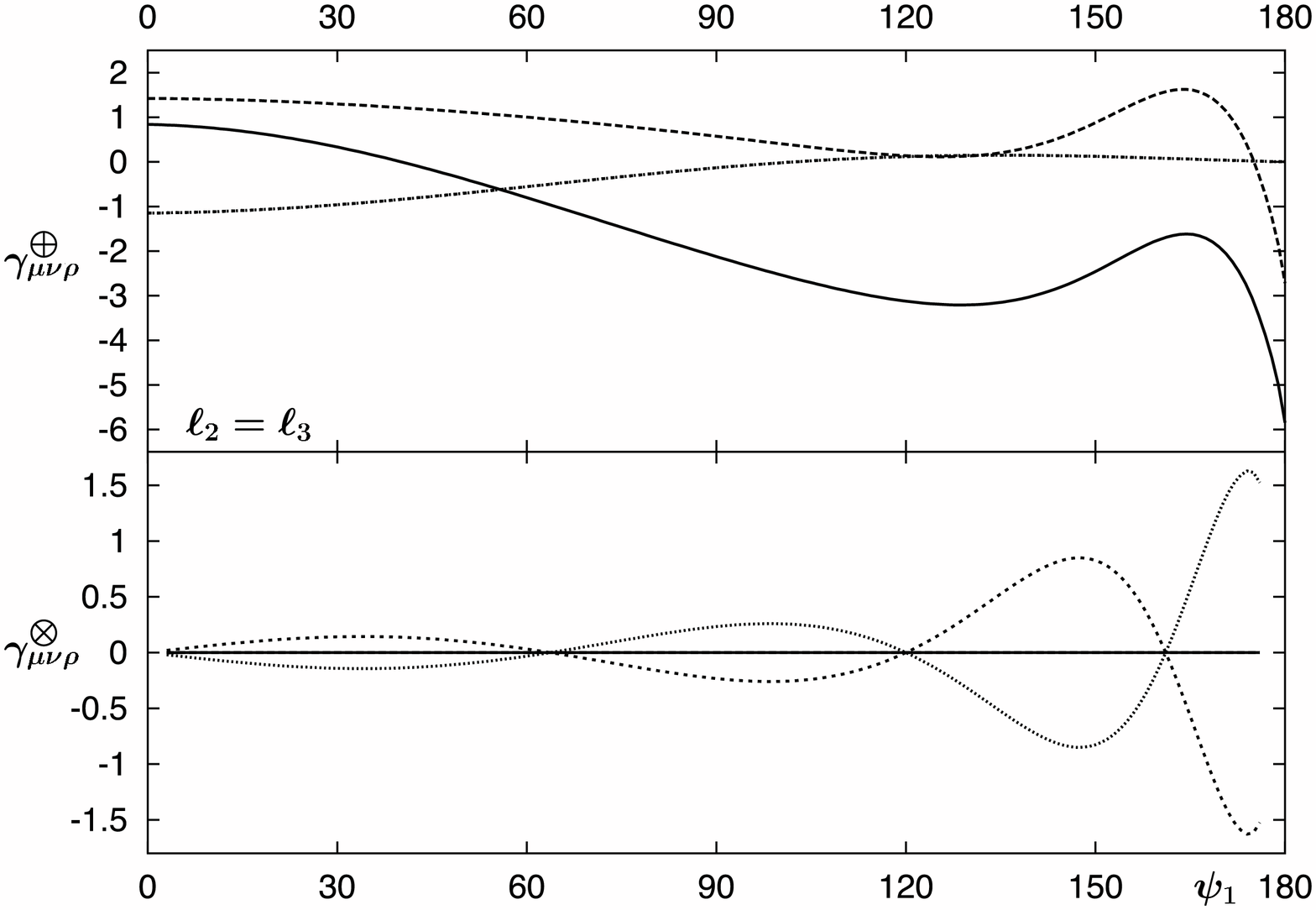}\\
\includegraphics[%
  bb=60bp 60bp 780bp 530bp,
  clip,
  width=1.0\columnwidth,
  keepaspectratio]{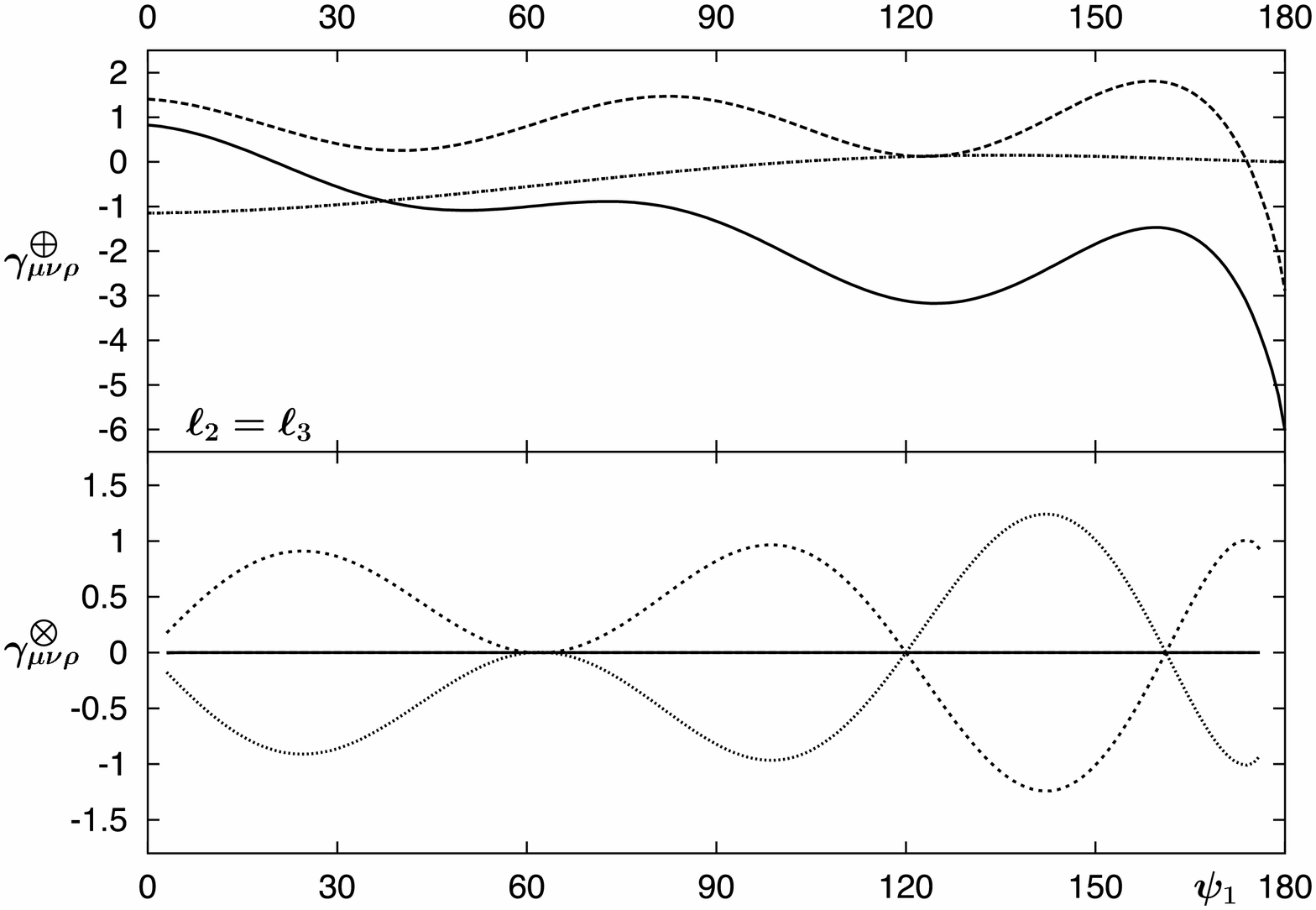} \tabularnewline
\end{tabular}

\caption{The top (bottom) panel shows the positive ($\gamma_{\mu\nu\rho}^{\oplus}$)
and negative ($\gamma_{\mu\nu\rho}^{\otimes}$) parity three-point
functions for the center of mass (orthocenter) projection convention.
We have assumed an isothermal sphere profile ($n=2$) and a fixed
ratio ($\ell_{2}/\ell_{3}=1$) and plotted as a function of the angle
$\psi_{1}$ (see Fig. \ref{cap:triangledef} for a definition).Line
styles for parity positive three-point functions are $\gamma_{+++}$
(solid), $\gamma_{+\times\times}$(long-dashed), $\gamma_{\times+\times}$
(dotted) and $\gamma_{\times\times+}$ (dashed) whereas for negative
parity $\gamma_{\times\times\times}$ (solid),$\gamma_{\times++}$
(long-dashed),$\gamma_{+\times+}$ (dotted) and $\gamma_{++\times}$
(dashed).\label{fig:gam:1}}
\end{figure}

\begin{figure}
\begin{tabular}{c}
 \includegraphics[
  bb=50bp 60bp 780bp 530bp,
  clip,
  width=1.0\columnwidth,
  keepaspectratio]{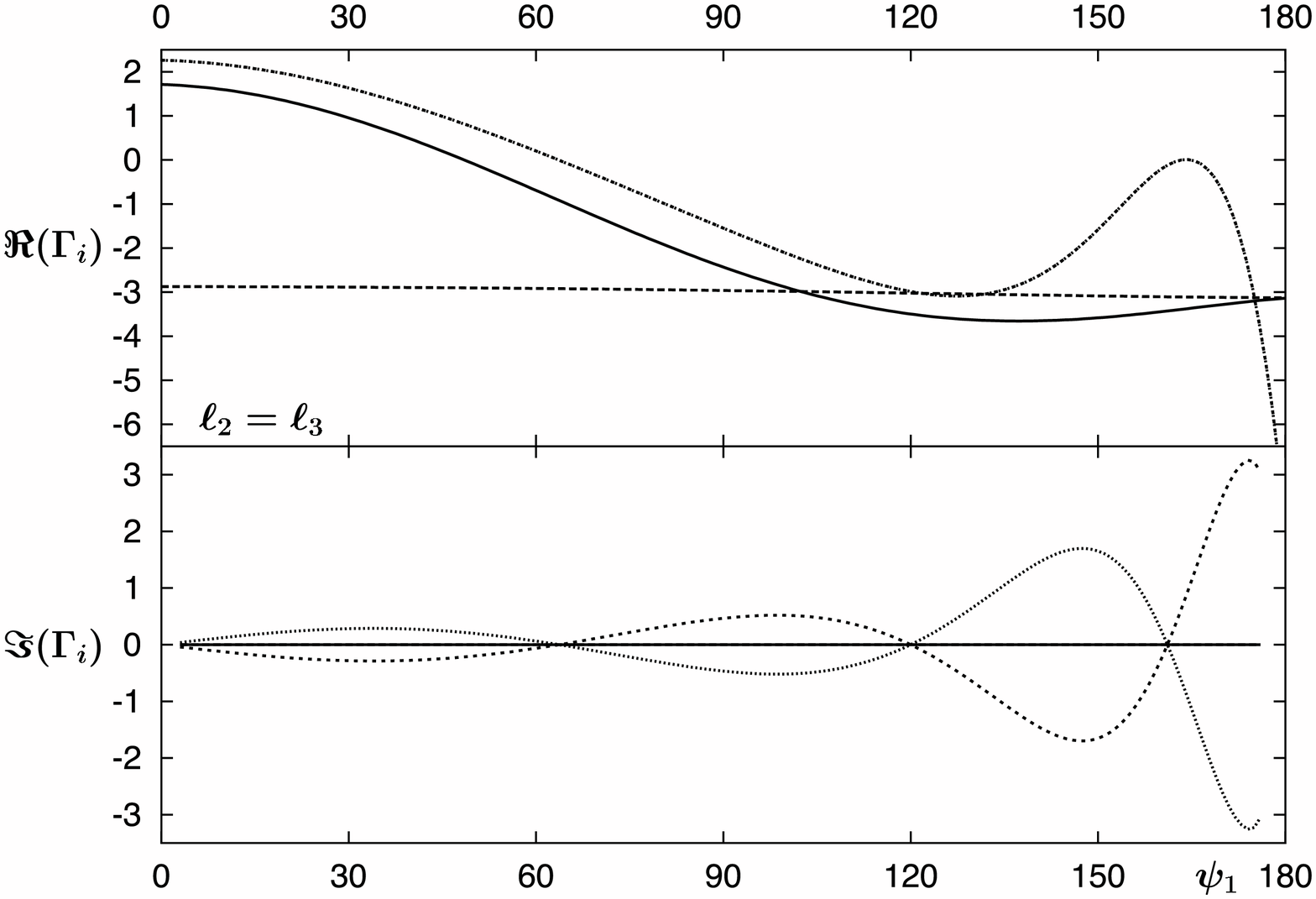}\\
\includegraphics[%
  bb=50bp 60bp 780bp 530bp,
  clip,
  width=1.0\columnwidth,
  keepaspectratio]{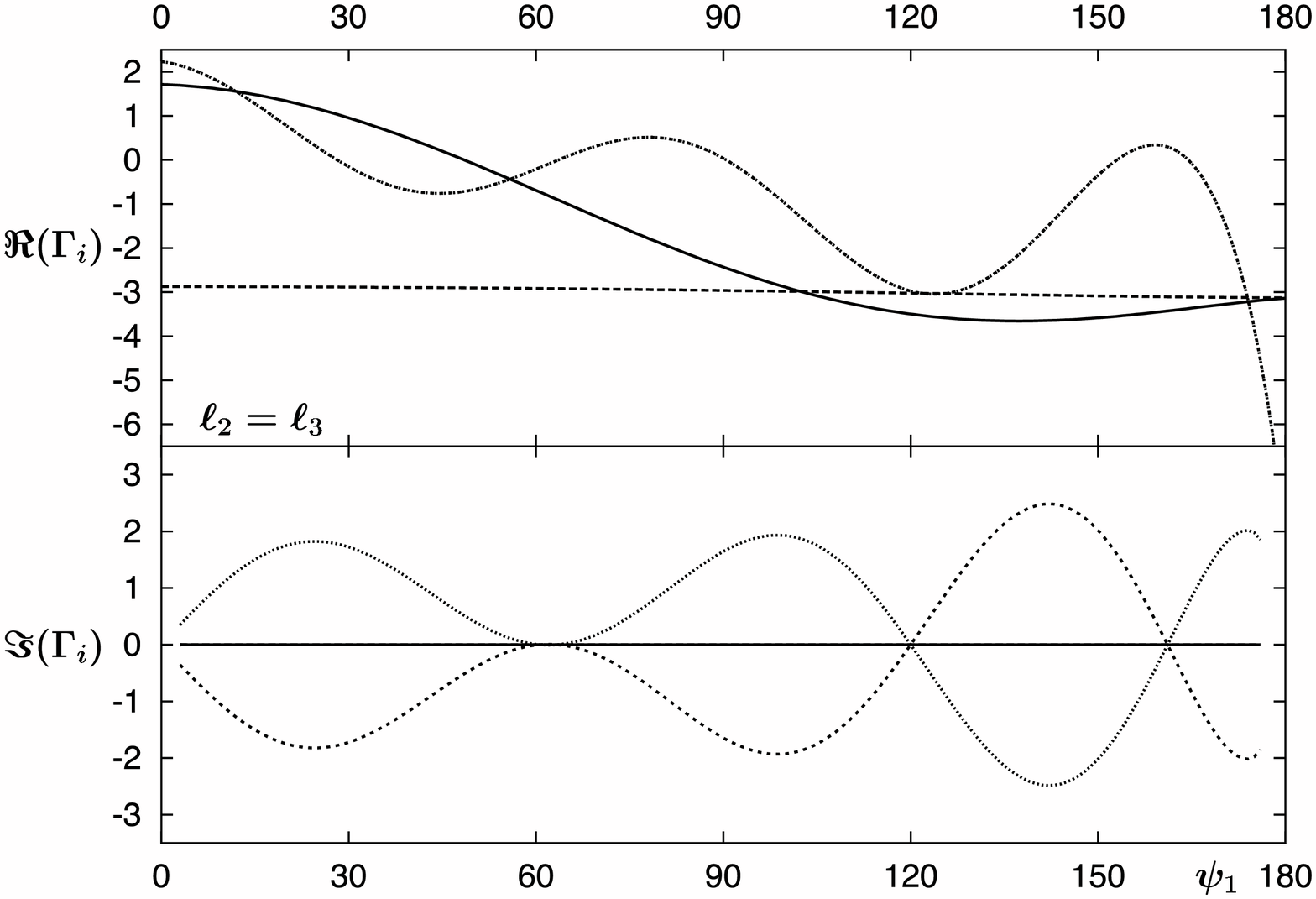} \tabularnewline
\end{tabular}

\caption{Same as Fig. \ref{fig:gam:1} but for the real (positive-parity)
and imaginary (negative parity) parts of $\Gamma_{i}$. Line styles
are as follows: $\Gamma_{0}$ (solid), $\Gamma_{1}$ (long-dashed),
$\Gamma_{2}$ (dotted), $\Gamma_{3}$ (dashed). \label{fig:Gam:1}}
\end{figure}

\begin{figure}
\begin{tabular}{c}
 \includegraphics[
  bb=50bp 60bp 780bp 530bp,
  clip,
  width=1.0\columnwidth,
  keepaspectratio]{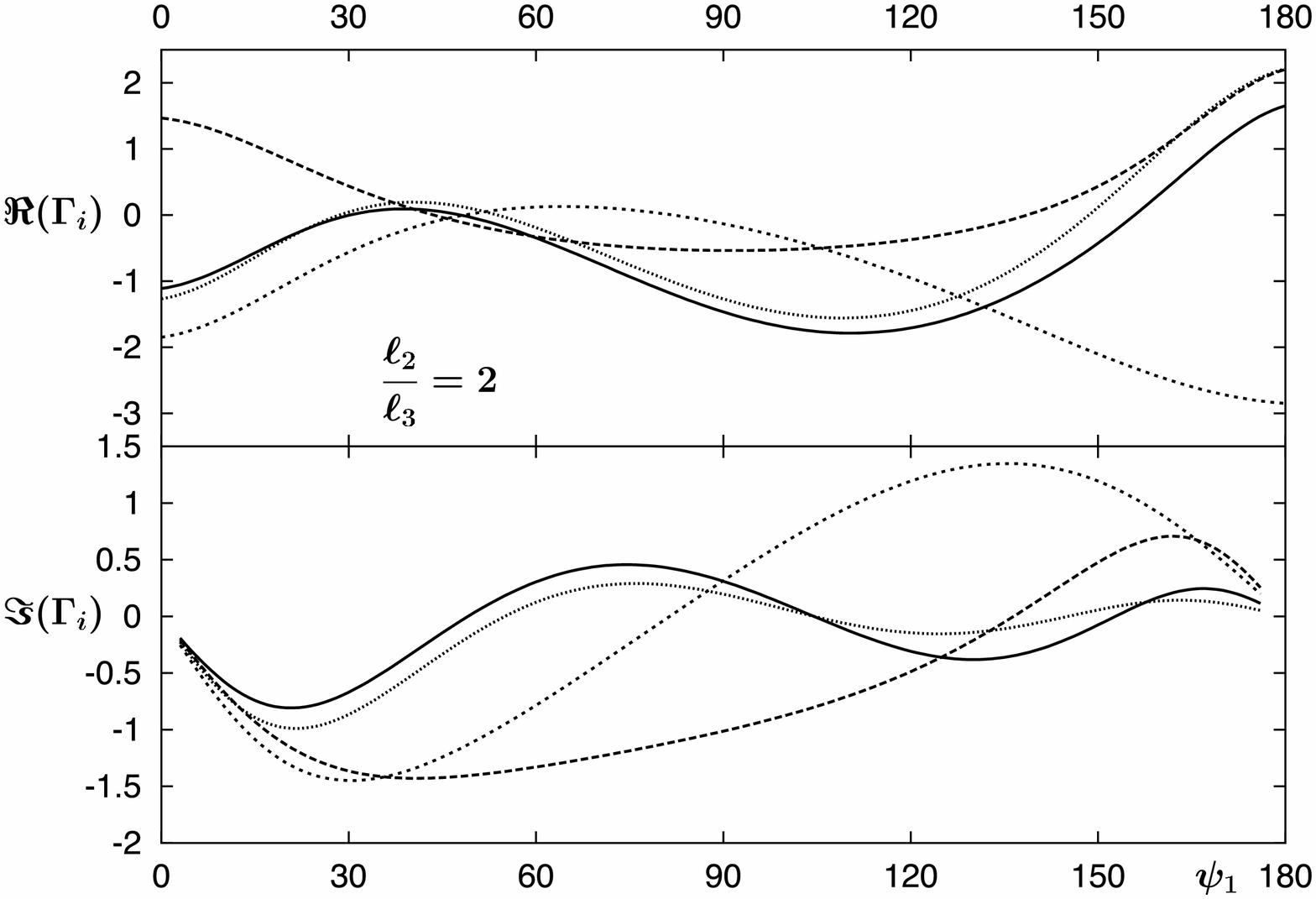}\\
\includegraphics[%
  bb=50bp 60bp 780bp 530bp,
  clip,
  width=1.0\columnwidth,
  keepaspectratio]{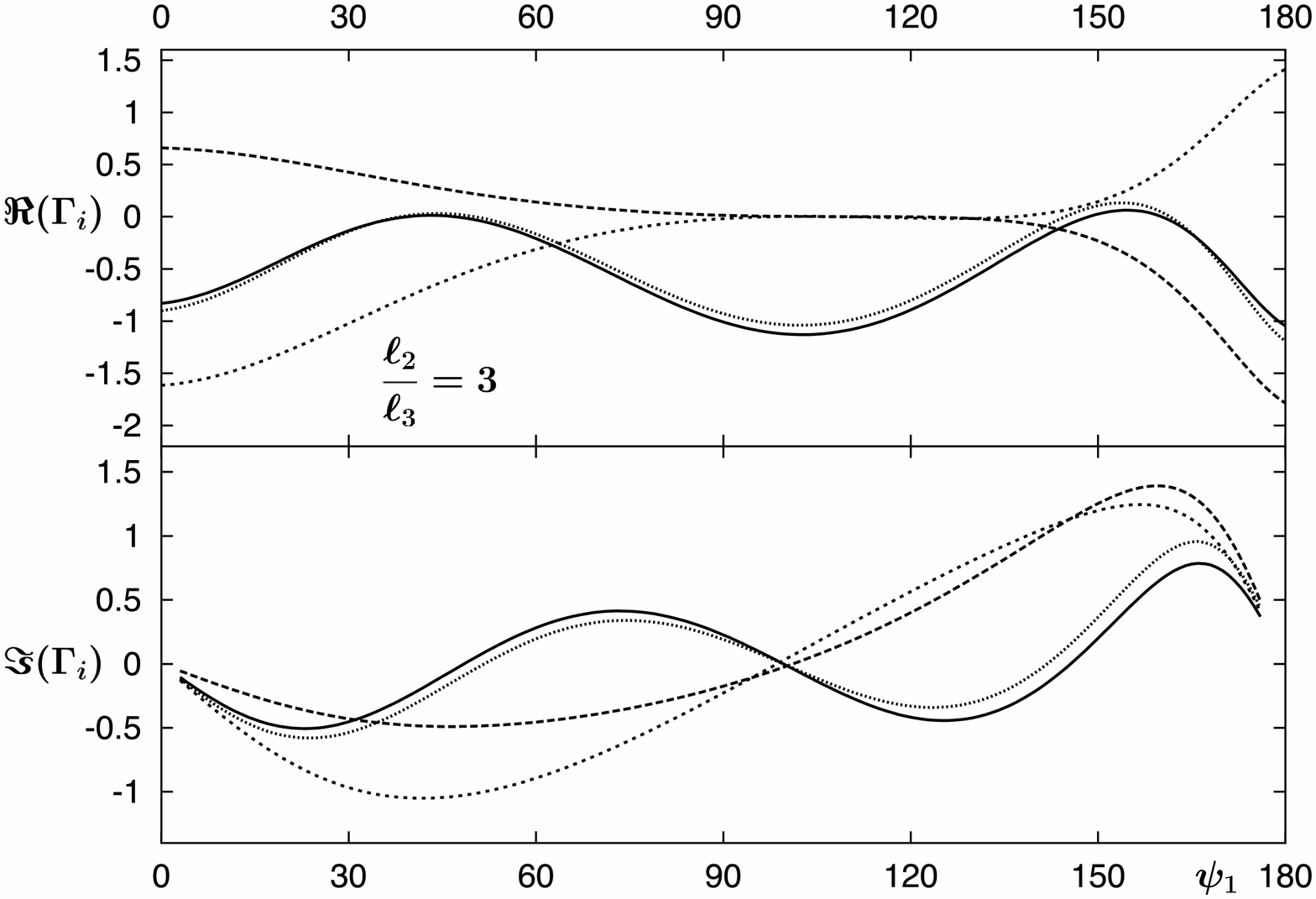} \tabularnewline
\end{tabular}

\caption{Same as Fig. \ref{fig:Gam:1} but only in the center of mass projection,
for sides ratio $\ell_{2}/\ell_{3}=2$ (top panel) and $\ell_{2}/\ell_{3}=3$
(bottom panel).\label{fig:Gam:2:3} }
\end{figure}

\begin{figure}
\begin{tabular}{c}
 \includegraphics[
  bb=30bp 60bp 770bp 550bp,
  clip,
  width=1.0\columnwidth,
  keepaspectratio]{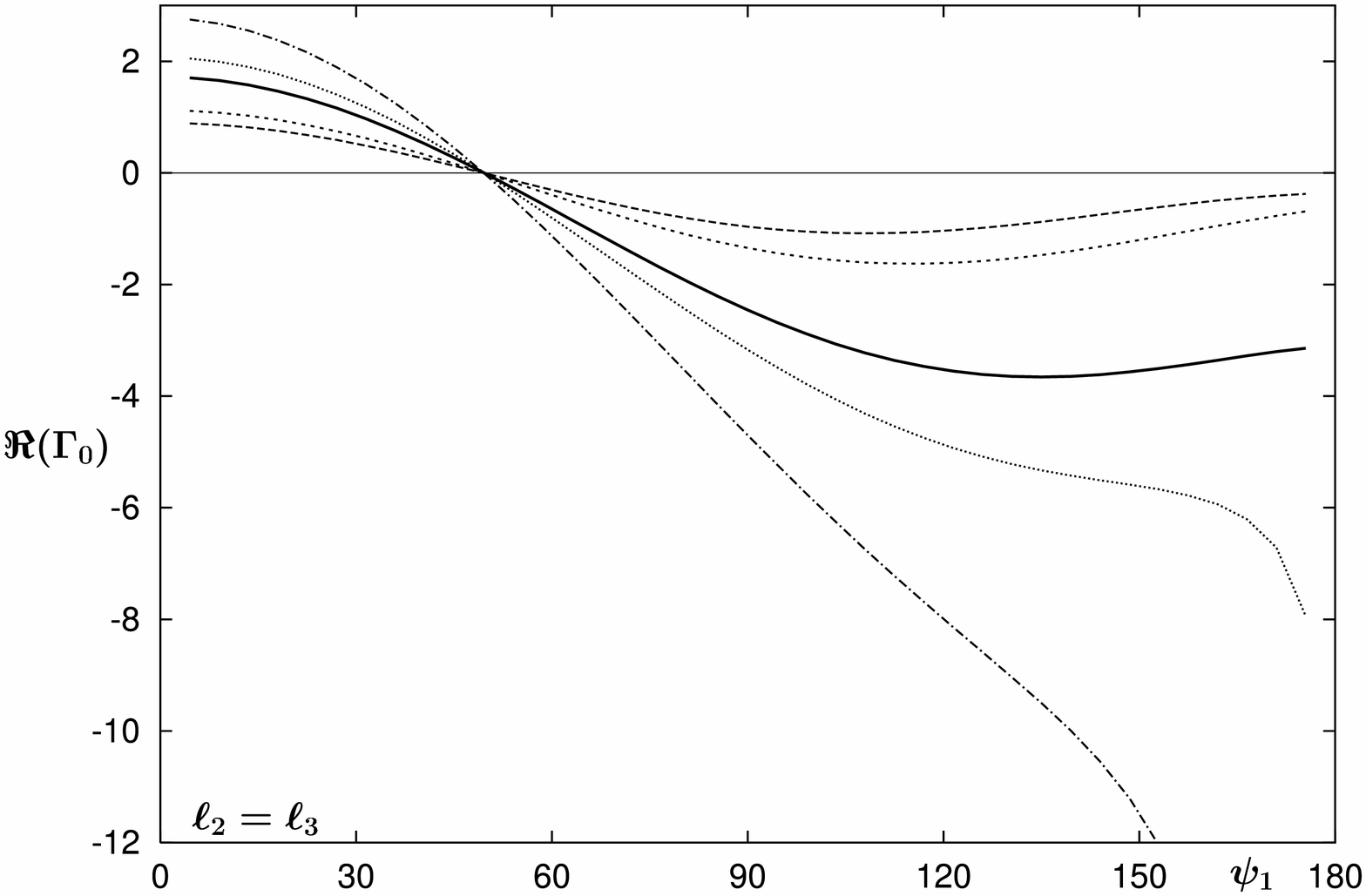}\\
\includegraphics[%
  bb=30bp 60bp 770bp 550bp,
  clip,
  width=1.0\columnwidth,
  keepaspectratio]{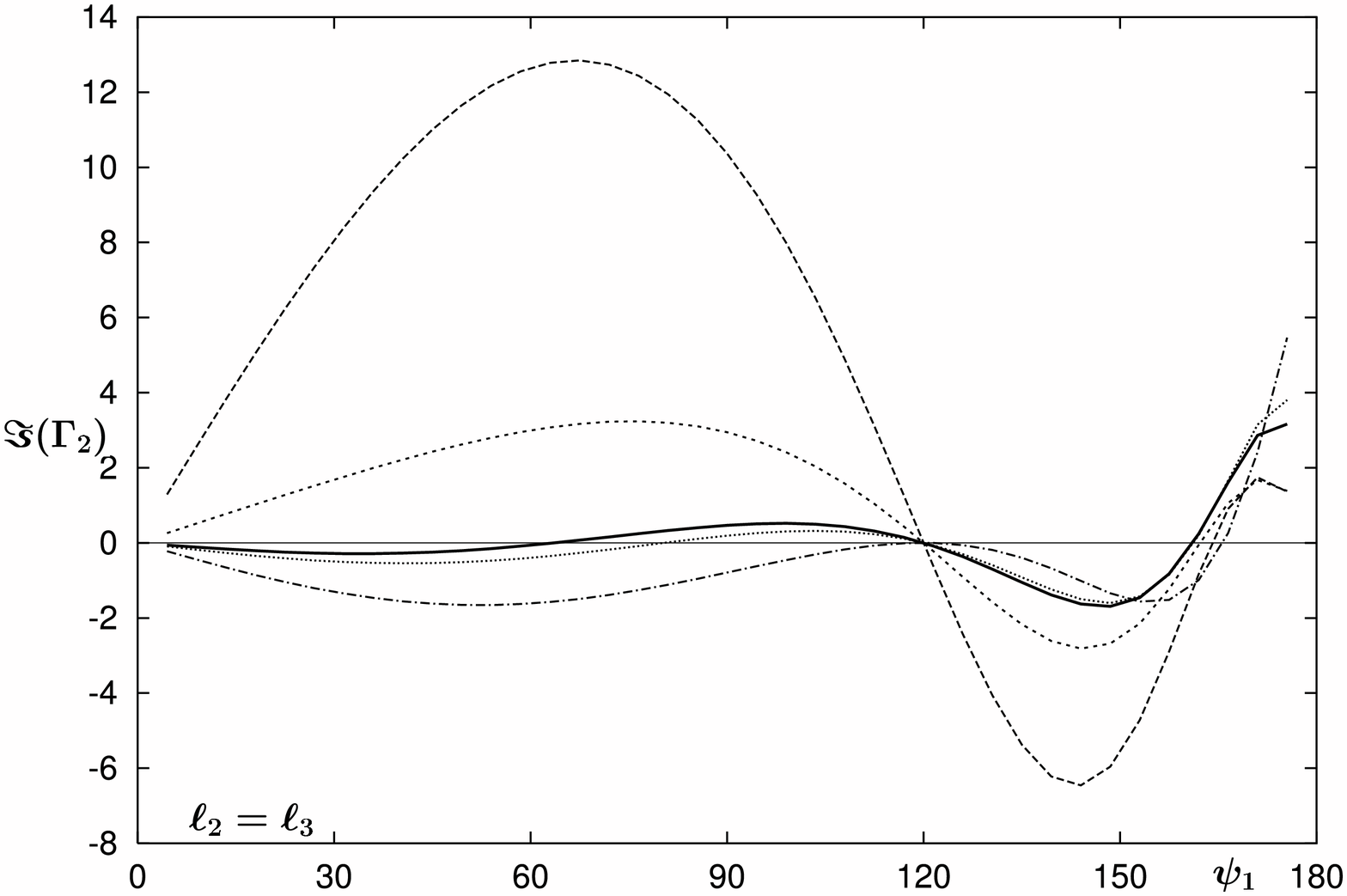} \tabularnewline
\end{tabular}

\caption{The real part of $\Gamma_{0}$ (top) and the imaginary part of $\Gamma_{2}$
(bottom) for different values of the profile index of the halo $n$
and $\ell_{2}=\ell_{3}$. Line styles are $n=1.2$ (long-dashed),
$n=2$ (solid), $n=2.1$ (dotted), $n=2.5$ (dot-dashed) \label{fig:Gam:nvar}}
\end{figure}

Figures \ref{fig:gam:1}, \ref{fig:Gam:1} and \ref{fig:Gam:2:3}
present our basic results for the isothermal profile case ($n=2$),
for fixed ratios of sides ($\ell_{2}/\ell_{3}=1,2,3$), as a function
of angle $\psi_{1}$ as defined in Fig. \ref{cap:triangledef}. Due
to the scaling in Eq. (\ref{eq:scaling}) a choice of sides ratio
completely describes other triangles with different overall scale
up to a normalization constant. Figure \ref{fig:gam:1} shows $\gamma_{\mu\nu\rho}$,
whereas Figs. \ref{fig:Gam:1} and \ref{fig:Gam:2:3} show $\Gamma_{i}$.
A comparison between different values of the power-law index $n$
is presented in Figs. \ref{fig:Gam:nvar} for $\Re(\Gamma_{0})$ and
$\Im(\Gamma_{2})$ when $\ell_{2}=\ell_{3}$.

Figures \ref{fig:gam:1} and \ref{fig:Gam:1} show results in two
different projection conventions. To go from the orthocenter to the
center of mass projection convention it is necessary to compute the
angles $\eta_{i}$ between the line joining the center of mass with
the $i^{\textrm{th}}$ vertex and the $\ell_{i}$ (see Fig. \ref{cap:triangledef})
and use the relations given by Eqs. (\ref{eq:Gamma0Rot}-\ref{eq:Gamma3Rot}).
Comparing both projections is useful to disentangle geometrical properties
from projection-dependent behavior. Comparing top and bottom panels
in Figs. \ref{fig:gam:1} and \ref{fig:Gam:1} we see that, qualitatively,
the orthocenter projection leads to {}``wigglier'' correlation functions.

Parity related features in the three-point functions for isoceles
triangles are evident: $\gamma_{\times+\times}=\gamma_{\times\times+}$
, $\gamma_{++\times}=-\gamma_{+\times+}$ and $\gamma_{\times\times\times}=\gamma_{\times++}=0$
in fig \ref{fig:gam:1} and $\Gamma_{2}=\Gamma_{3}^{*}$ and $\Im(\Gamma_{0})=\Im(\Gamma_{1})=0$
in fig \ref{fig:Gam:1}. Furthermore, for equilateral triangles $\gamma_{+\times\times}=\gamma_{\times+\times}=\gamma_{\times++}$
and $\gamma_{\times++}=\gamma_{+\times+}=\gamma_{++\times}=0$. Other
features that these figures show appear more difficult to predict;
for example, the local extrema of $\gamma_{+++}$ for configurations
close to equilateral triangles are not exactly located at $2\pi/3$
and depend on the projection convention (after all, the average over
the position of the halo does depend on projection convention). In
addition, points where some $\gamma_{\mu\nu\rho}$ are equal to each
other change location (and can disappear or appear) as projection
convention is changed.

Figure \ref{fig:Gam:2:3} shows what happens as we consider triangles
other than isosceles. Note that now all the parity-negative three-point
functions are non-zero. The fact that $\Gamma_{1}\sim-\Gamma_{2}^{*}$
(and accordingly that $\Gamma_{0}\sim\Gamma_{3}$) can be understood
as follows. Indeed, when $\cos(\pi-\psi_{1})=1/4$ ($1/6$ for bottom
panel) the configuration is isoceles in $\theta_{3}$, thus ensuring
that $\Gamma_{1}=\Gamma_{2}^{*}$ by parity. Around this angle, it
follows from parity that $\Gamma_{2}(\psi_{1}+\epsilon,\ell_{2},\ell_{3})\sim\Gamma_{1}^{*}(\psi_{1}-\epsilon,\ell_{2},\ell_{3})$
for $\epsilon\ll\ell_{2}/\ell_{3}$. In addition, as $\ell_{2}/\ell_{3}$
increases, the product $\gamma_{1}\gamma_{2}$ will start to dominate
the three-point function, and thus parity properties become essentially
those of the two-point function, insuring that $\Gamma_{1}\sim-\Gamma_{2}^{*}$.
These arguments explain why as $\ell_{2}/\ell_{3}$ increases (compare
top and bottom panels in Fig. \ref{fig:Gam:2:3}) $\Gamma_{1}$ gets
closer to $-\Gamma_{2}^{*}$ and $\Gamma_{0}$ to $\Gamma_{3}$.

Figure \ref{fig:Gam:nvar} shows how the correlation function $\Re(\Gamma_{0})$
and $\Im(\Gamma_{2})$ depend on the slope of the profile $n$. The
zero crossing of $\Im(\Gamma_{2})$ at $2\pi/3$ can be explained
by parity; however, the other zero crossings (and the number of them)
depend on the profile slope $n$. For $\Re(\Gamma_{0})$ there is
a zero crossing at $\psi_{1}\sim50^{o}$, which appears robust to
changes in $n$, but we found no simple explanation for this.

\begin{figure}
 \includegraphics[%
  bb=50bp 60bp 780bp 530bp,
  clip,
  width=1.0\columnwidth,
  keepaspectratio]{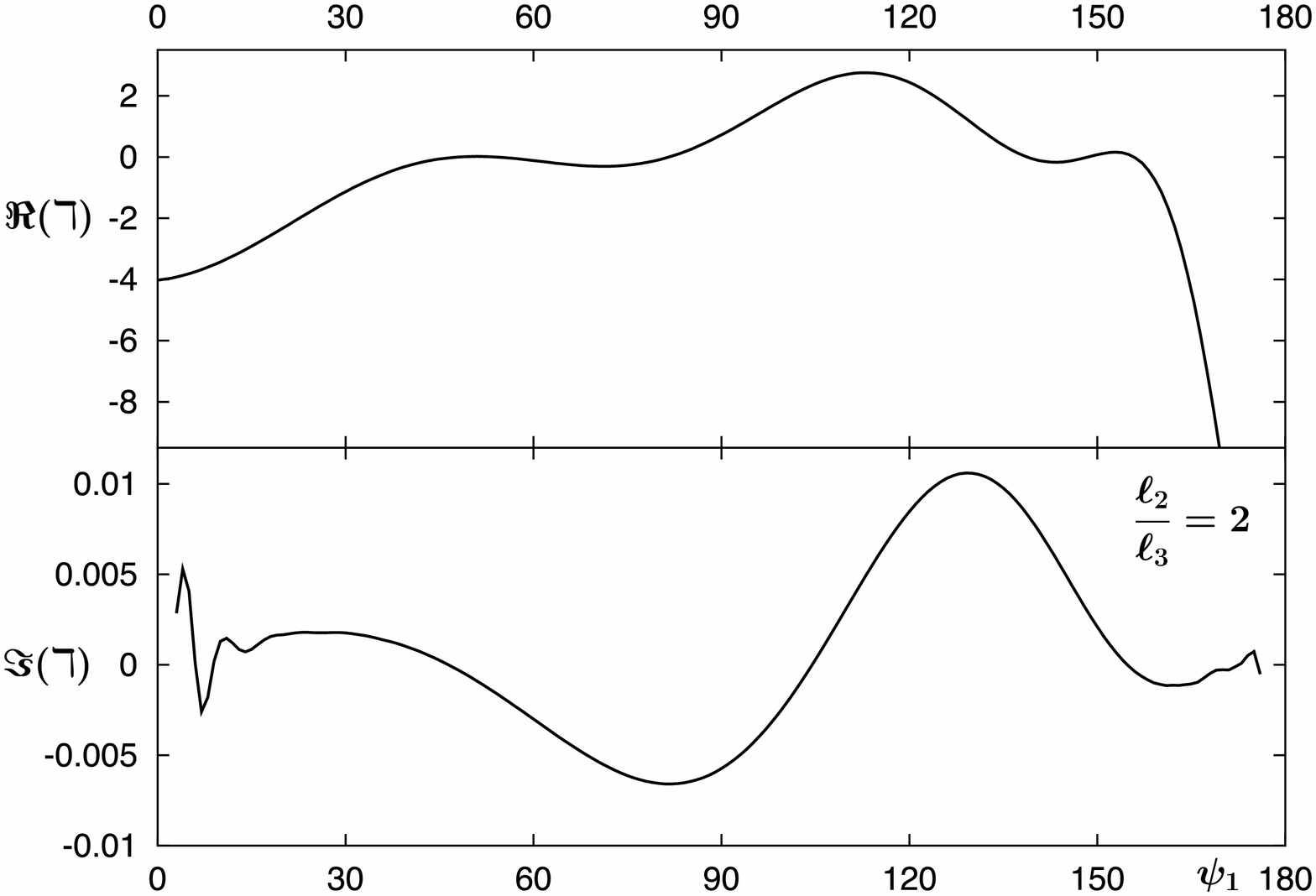}

\caption{\label{fig:tau:etrange}The real (top) and imaginary part of $\daleth$
for the $\ell_{2}/\ell_{3}=2$ case as a function of $\psi_{1}$.
The non-zero imaginary part shows that there is no preferred projection
convention where all negative-parity three-point functions vanish.}
\end{figure}

We now consider again the question raised at the end of section \ref{sub:3pt:def}
regarding the existence of a preferred projection, in the context
of our simple model. The condition in Eq. (\ref{eq:simpleprojectionncondition})
can be rewritten as \begin{equation}
\daleth=\Gamma_{0}\left(\Gamma_{1}\Gamma_{2}\Gamma_{3}\right)^{*}\epsilon~\mathbb{R}.\label{eq:conditionquitue}\end{equation}
 Figure \ref{fig:tau:etrange} shows $\Re(\daleth)$ and $\Im(\daleth)$
for triangle with $\ell_{2}/\ell_{3}=2$. We see from the bottom panel
that the imaginary part does not vanish, thus it is not possible to
find a preferred projection where all negative-parity three-point
functions vanish {[}but of course we can make three of them vanish
by using Eqs. (\ref{eq:Gamma0Rot}-\ref{eq:Gamma3Rot}){]}. Note however,
that the relation in Eq. (\ref{eq:conditionquitue}) is close to being
satisfied, at least approximately, for almost all $\psi_{1}$. %

\section{The shear three-point functions in numerical simulations\label{sec:cinq}}

We now describe the results of measuring the shear three-point function
in N-Body simulation. For details about the simulations and the procedure
we followed to make the measurements see Appendix \ref{sec:Nbody}.
Basically, we created shear maps with three different resolutions
that cover a large range of scales (roughly from 4 arc seconds up
to 1, 10 and 40 arc minutes). In practice, we measured the $\gamma_{\mu\nu\rho}$
in the center of mass projection convention, and then transformed
to the $\Gamma_{i}$.

\subsection{Results}

\begin{figure}
\begin{center}\includegraphics[%
  height=1.0\columnwidth,
  keepaspectratio,
  angle=270]{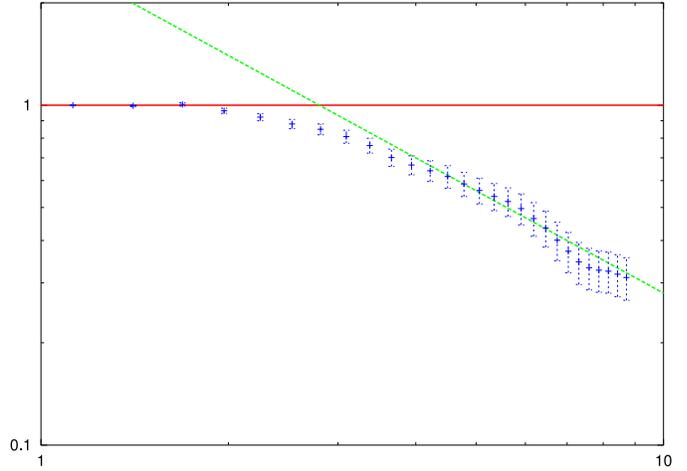}\end{center}

\caption{Scaling test for $\ell_{0}=1.125'$. The points with error bars represent
$S=\bar{\ell\Gamma_{1}}(\ell)/(\ell_{0}\bar{\Gamma_{1}}(\ell_{0}))$.
The solid line is $S=1$ ($n=2$), whereas the dotted line denotes
to which according to Eq. (\ref{eq:scaling}) correspond to $n\simeq\approx2.3$.\label{fig:scale} }
\end{figure}

We start by comparing the scaling of the three-point functions to
see, using Eq. (\ref{eq:scaling}), whether the effective value of
the profile index is reasonable compared to expectations based on
dark matter halo profiles such as Eq. (\ref{eq:profile:gen}). Figure
\ref{fig:scale} shows results from the medium resolution measurements,
where we have scaled $\Re(\Gamma_{1})$ assuming an $n=2$ profile.
More precisely, given an isosceles triangle where two sides are equal
to $\ell$, we have fitted for $\bar{\Gamma_{1}}$ and $\alpha(\ell)$
so that\begin{equation}
\Re(\Gamma_{1}(\psi_{1},\ell,\ell))\sim\bar{\Gamma_{1}}(\ell)(1+\alpha(\ell)\psi_{1}),\end{equation}
 where $\alpha(\ell)$ incorporates the dependence on the angle between
the two equal sides. In practice we found that variations of $\alpha$
with $\ell$ are small, thus they have been neglected. To avoid artificial
deviations from scaling due to our binning (see discussion below),
we have kept our triangles far enough from being collapsed, i.e. we
restrict $\pi/9\leq\psi_{1}\leq8\pi/9$. The results in Fig.~\ref{fig:scale}
show that the scaling between~$1$ and $3$ arc-min is consistent
with that of an isothermal sphere, whereas for larger angular scales
$\ell\bar{\Gamma_{1}}(\ell)\sim\ell^{-1}$ and thus the effective
profile index increases toward $n=7/3\sim2.3$ {[}see Eq. (\ref{eq:scaling}){]}.
This is consistent with the hypothesis that we are only sensitive
to the index of halos that contribute the most at the scale we are
probing. Indeed, at the maximum of the lensing window function, $z\sim0.4$,
masses in the range $10^{13}M_{\odot}$ to $10^{15}M_{\odot}$ have
$r_{0}$ {[}where $n_{\textrm{eff}}=2,$ see Eq. (\ref{eq:profile:gen}){]}
of the order of a few arc-minutes.

\begin{figure}
\begin{tabular}{c}
 \includegraphics[
  bb=50bp 80bp 770bp 560bp,
  clip,
  width=1.0\columnwidth]{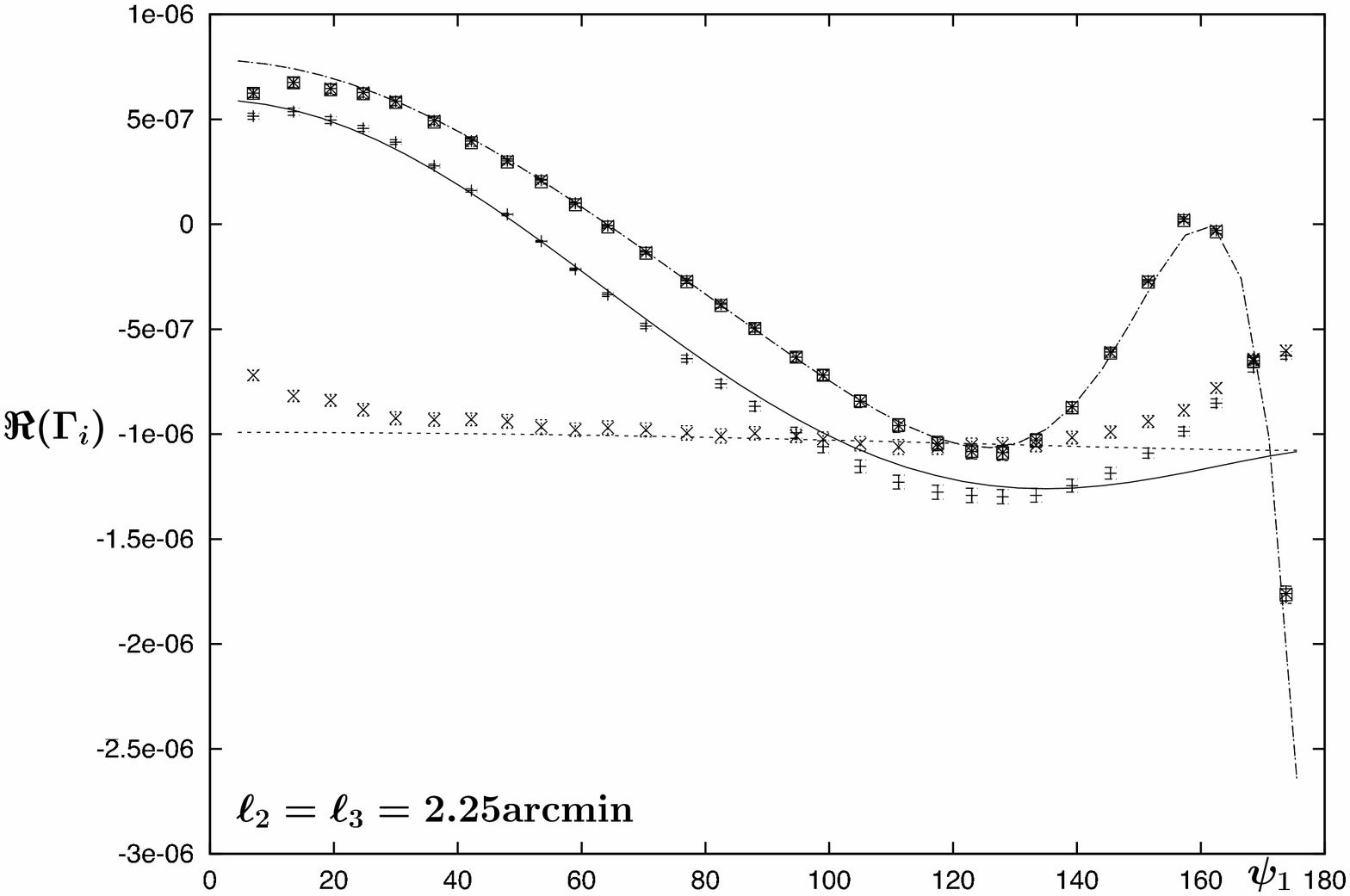}\\
\includegraphics[%
  bb=50bp 80bp 770bp 560bp,
  clip,
  width=1.0\columnwidth]{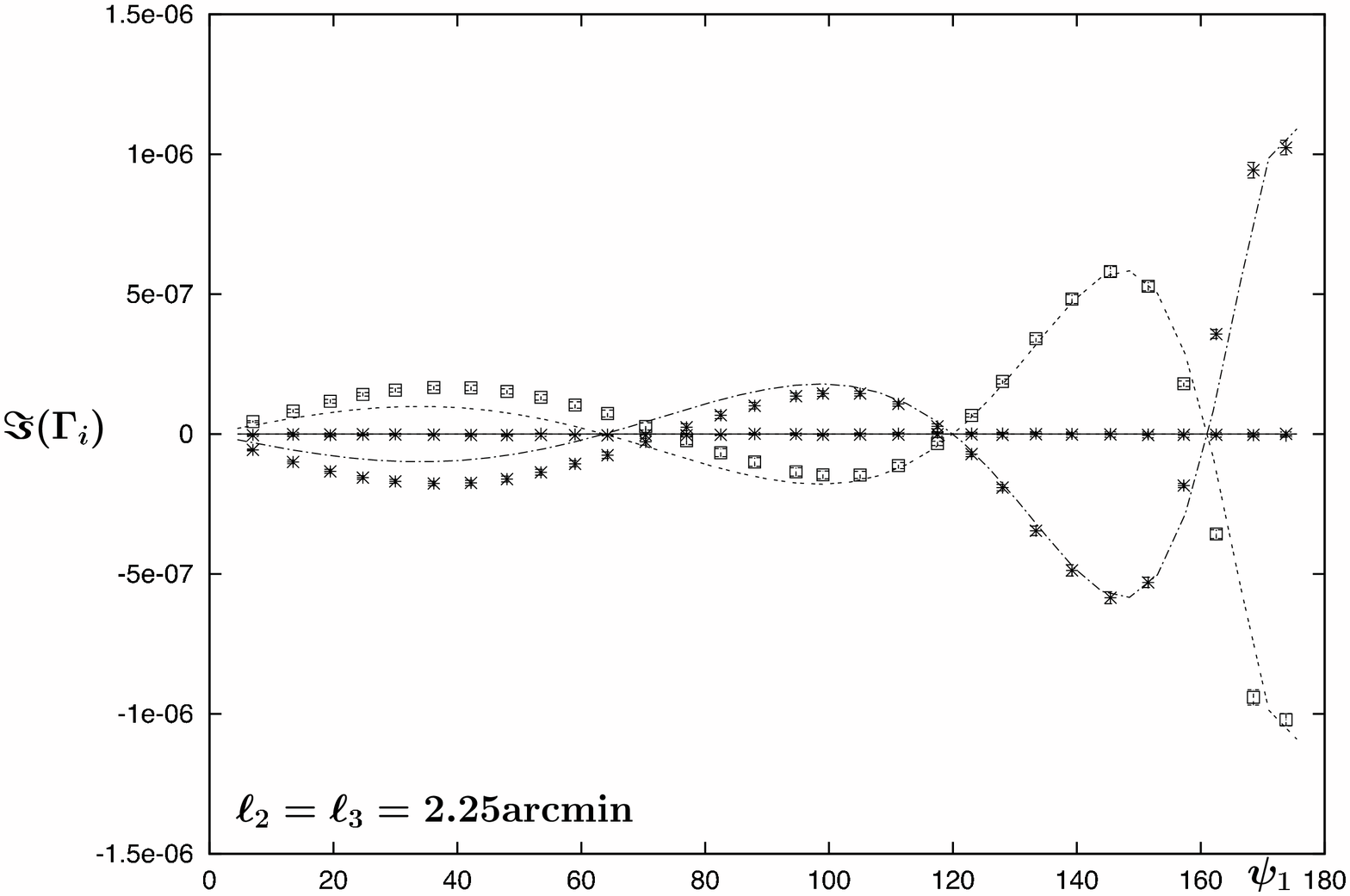} \tabularnewline
\end{tabular}

\caption{Predictions from our $n=2$ model compared against measurements in
simulations for isosceles triangles with $\ell=2.25$ arc minutes.
The top (bottom) panel shows the real (imaginary) part of $\Gamma_{1}$,
and the model arbitrary amplitude has been adjusted to match $\Gamma_{2}$.\label{fig:mauvaisaccord} }
\end{figure}

Given the results in Fig.~\ref{fig:scale}, we can now compare the
$n=2$ model to our measurements in simulations at arc minute scales
to check whether they agree (up to an arbitrary constant that our
model does not predict). Figure \ref{fig:mauvaisaccord} shows the
comparison between the $n=2$ model (with amplitude fixed by maximizing
agreement with $\Gamma_{2}$) and simulations for triangles with $\ell_{2}=\ell_{3}=2.25$
arc-minutes. We see that by adjusting a single amplitude, all other
$\Gamma_{i}$ show good agreement as well, giving support to our simple
model. This result is not surprising given that an isothermal profile
was shown to agree with a calculation based on the halo model in \cite{2003ApJ...584..559Z},
and the latter was found to be in good agreement with measurements
in numerical simulations in \cite{2003ApJ...583L..49T,2003astro.ph..4034T}.

\begin{figure}
\begin{tabular}{c}
 \includegraphics[
  bb=50bp 80bp 770bp 560bp,
  clip,
  width=1.0\columnwidth]{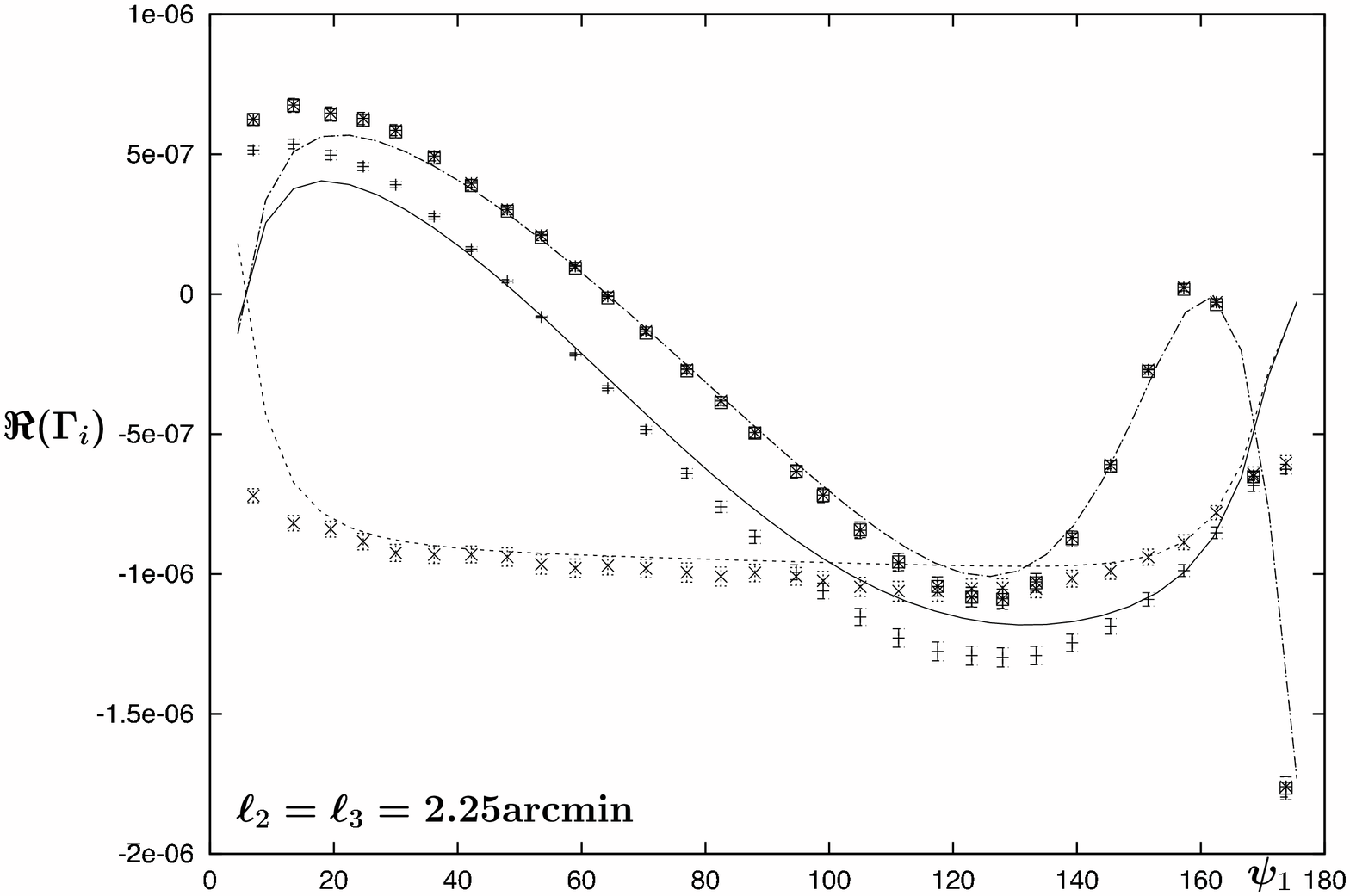}\\
\includegraphics[%
  bb=50bp 80bp 770bp 560bp,
  clip,
  width=1.0\columnwidth]{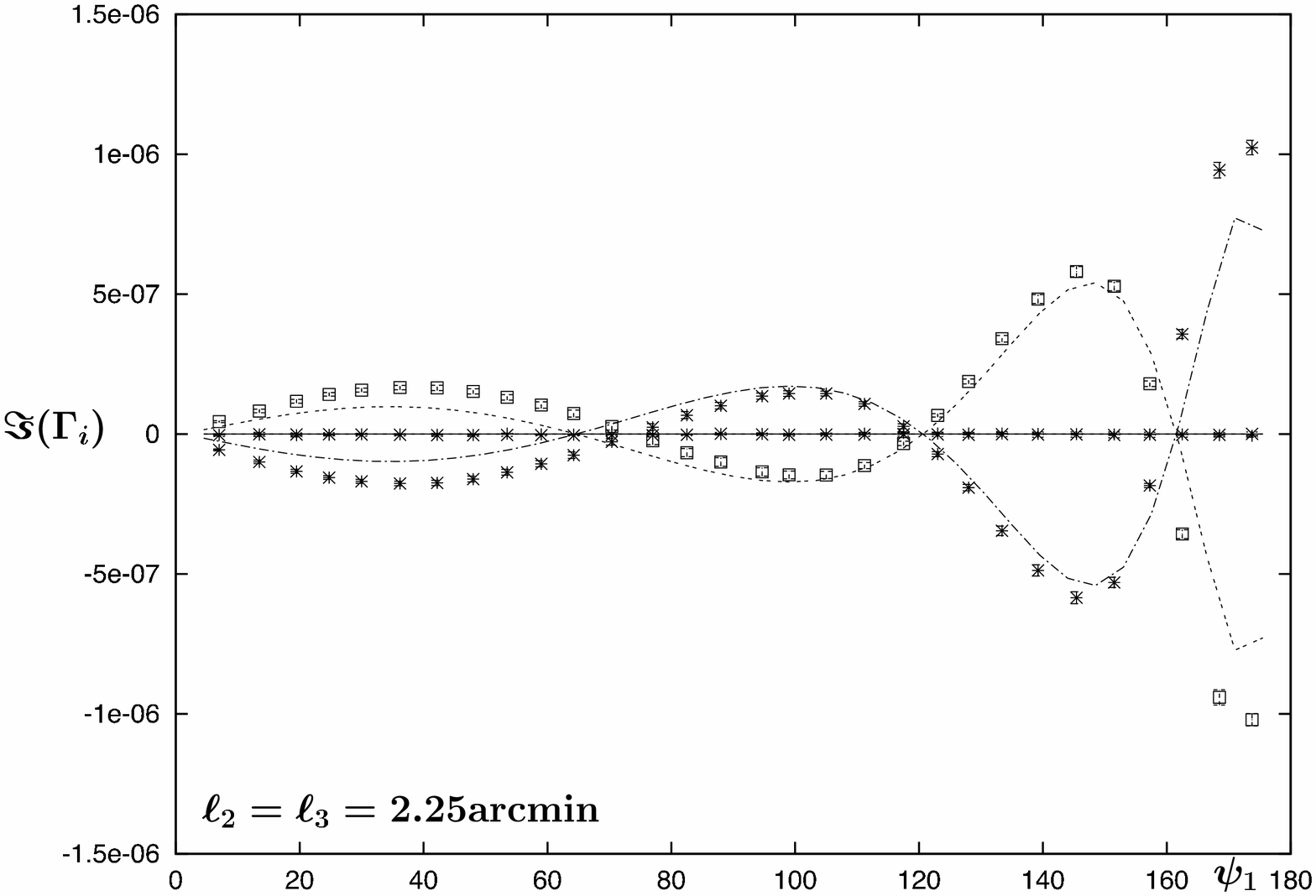} \tabularnewline
\end{tabular}

\caption{Same as previous Figure, but modifying the model prediction to mimic
the effect of binning in the simulations, approximately, as in Eq.
(\ref{eq:modif:gamma}).\label{fig:accordbinning}}
\end{figure}

It is apparent from Fig. \ref{fig:mauvaisaccord} that as the triangle
becomes collapsed ($\psi_{1}\rightarrow0^{o},180^{o}$) the agreement
with the model is not as good, particularly in the parity positive
case. This can be understood as a result of the effect of binning
in the simulation measurements, where many configurations contribute
to each bin. In the case we are considering here, it amounts to an
error of order 10\% on the length of $\ell_{2}$ and $\ell_{3}$ and
about 2\% on the angle $\psi_{1}$. On can estimate the effect of
binning by computing in our model $\tilde{\Gamma_{i}}$ defined as
instead \begin{equation}
\tilde{\Gamma}_{i}=\frac{\Gamma_{i}(\psi_{1},(1+\epsilon)\ell,\ell)+\Gamma_{i}(\psi_{1},\ell,(1+\epsilon)\ell)}{2},\label{eq:modif:gamma}\end{equation}
 with $\epsilon=\Delta\ell/\ell\sim0.1$. This prediction is compared
to the same measurements in Fig. \ref{fig:accordbinning}. The results
show that the behavior near collapsed triangles, such as the sharp
divergence of $\Re(\Gamma_{1})$, can be explained by the discreteness
error due to binning. Generically, our model shows that the effect
of a small error in the determination of the length of the triangle
sides can translate into significant corrections to the three-point
functions for nearly collapsed triangles. Another noticeable discrepancy
between our model and the numerical simulations are the slightly displaced
zeros of $\Im(\Gamma_{2})$ and $\Im(\Gamma_{3})$ and their amplitudes
for $\psi_{1}<60^{o}$. The former can be improved by slightly changing
the profile index, as shown in Fig. \ref{fig:Gam:nvar} the first
zero in these three-point functions is very sensitive to small variations
in the profile index. The difference in amplitude can also be seen
in similar plots in \cite{2003ApJ...583L..49T} when using the full
halo model; this suggests perhaps that such deviations could be due
to the assumption of spherical halos. It is worth exploring this issues
further as they may provide a novel way of testing dark matter halo
profiles and shapes. Some steps in this direction have been already
taken in \cite{2003astro.ph..4034T}, where the two- and three-point
function of the convergence field were used to constrain parameters
of halos such as concentration and inner profile slope.

\begin{figure}
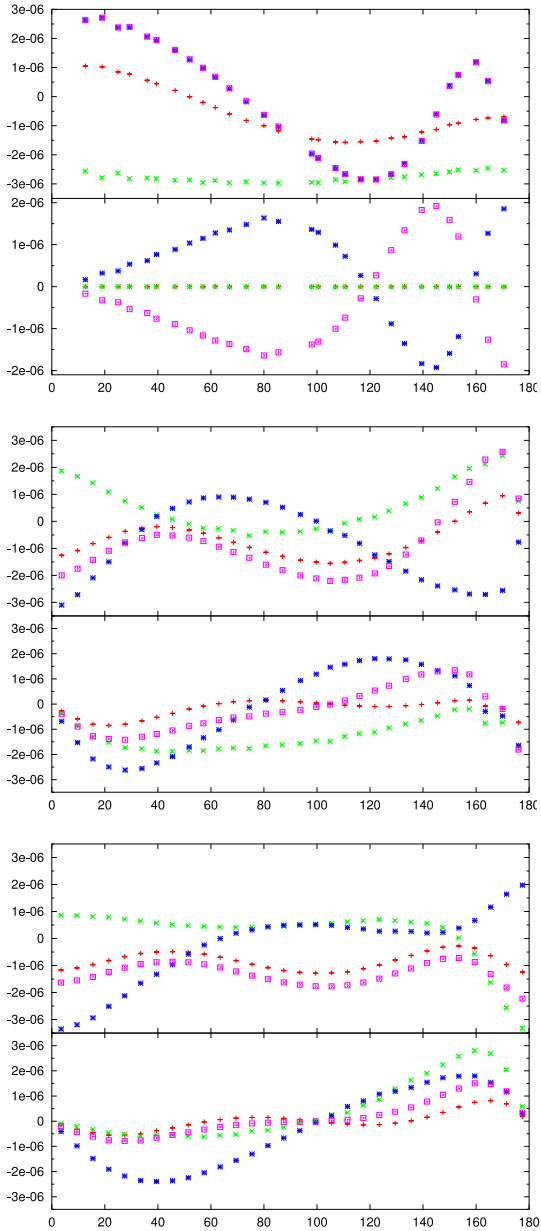

\noindent \begin{center}\begin{tabular}{c}
\includegraphics[%
  bb=55bp 80bp 550bp 750bp,
  clip,
  width=0.60\columnwidth,
  keepaspectratio,
  angle=270]{Figs/Gammas_6_6_2048.epsi}\tabularnewline
\includegraphics[%
  bb=55bp 80bp 550bp 750bp,
  clip,
  width=0.60\columnwidth,
  keepaspectratio,
  angle=270]{Figs/Gammas_12_6_2048.epsi}\tabularnewline
\includegraphics[%
  bb=55bp 80bp 550bp 750bp,
  width=0.60\columnwidth,
  keepaspectratio,
  angle=270]{Figs/Gammas_15_5_2048.epsi}\tabularnewline
\end{tabular} \end{center}

\caption{$\Gamma_{i}$ measured for $\ell_{2}=0.42'$. In each panel top (bottom)
part shows positive (negative) parity three-point functions. Panels
correspond to isoceles triangles (top), $\ell{}_{2}=2\ell_{3}$ (middle)
and $\ell_{2}=3\ell_{3}$ (bottom). Symbols are as follows $\Gamma_{0}$
(plus),$\Gamma_{1}$ (cross), $\Gamma_{2}$ (star) and $\Gamma_{3}$
(box).\label{fig:Gam:2028:6:1} }
\end{figure}

\begin{figure}
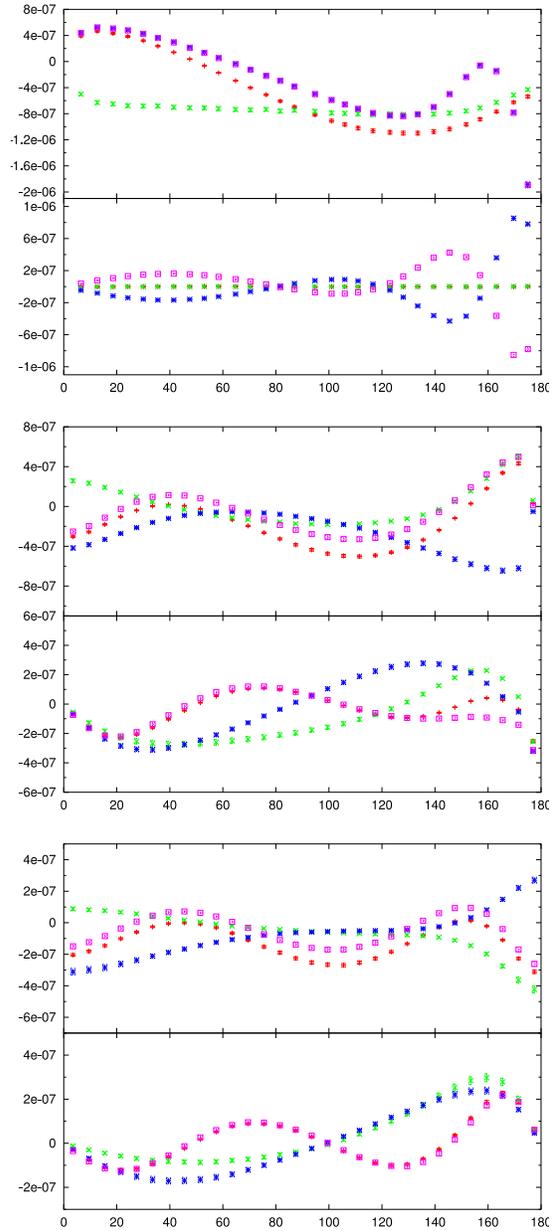

\noindent \begin{center}\begin{tabular}{c}
\includegraphics[%
  bb=55bp 80bp 550bp 750bp,
  clip,
  width=0.60\columnwidth,
  keepaspectratio,
  angle=270]{Figs/Gammas_10_10_512.epsi}\tabularnewline
\includegraphics[%
  bb=55bp 80bp 550bp 750bp,
  clip,
  width=0.60\columnwidth,
  keepaspectratio,
  angle=270]{Figs/Gammas_20_10_512.epsi}\tabularnewline
\includegraphics[%
  bb=55bp 80bp 550bp 750bp,
  clip,
  width=0.60\columnwidth,
  keepaspectratio,
  angle=270]{Figs/Gammas_30_10_512.epsi}\tabularnewline
\end{tabular} \end{center}

\caption{Same as previous Figure for $\ell_{2}=2.8'$. \label{fig:Gam:512:10:1}}
\end{figure}

Finally, we present results from simulations for isosceles and non-isosceles
configurations, from our high resolution measurements, at $\ell_{2}=0.42'$
(Fig. \ref{fig:Gam:2028:6:1}) and middle resolution one at $\ell_{2}=2.8'$
(Fig. \ref{fig:Gam:512:10:1}). These should be compared with the
top panel in Fig. \ref{fig:Gam:1} for the isosceles case, and Fig.
\ref{fig:Gam:2:3} for non-isosceles triangles. We can see that for
$\ell_{2}=2.8'$ the agreement with Figs. \ref{fig:Gam:1}-\ref{fig:Gam:2:3},
as expected from the scaling test that suggests that the effective
index is close to n = 2. However, for $\ell_{2}=0.42'$ we see that
there are significant differences, the three-point functions changed
as expected from the results of our analytical model in Fig. \ref{fig:Gam:nvar}
when the profile index becomes smaller than $n=2$.

We have also made measurements in lower resolution sets (not presented
here) that allows us to probe larger scales. Again we see consistent
results with those expected from Fig. \ref{fig:Gam:nvar} for indices
$n>2$, in particular regarding the dependence of number of zeros
as the scale is changed. However, one cannot extend this study to
significantly larger scales, as contributions from more than a single
halo become important and our simple model breaks down.

\section{Estimators of cosmic shear three-point functions\label{sec:estim}}

We now turn to the problem of building a simpler estimator of the
shear three-point function. One obvious solution is to combine them
to reconstruct the weak lensing convergence bispectrum \cite{2005A&A...431....9S}.
However, equations involved into the computation of the bispectrum
from the shear three-point functions are very difficult to compute
numerically. They correspond to the inversion of a non-local equation;
a task that usually cannot be fulfilled from real data without some
regularization of the problem.

The so-called $M_{ap}$ statistic filters provide such regularization.
They have been used successfully to reconstruct the filtered convergence
two-point function from measurement of the shear two-point functions
\cite{2002AA...396....1S}. Another great advantage of these family
of compensated filters is that, to some extent, they can be custom
designed to have a finite real space support (or at least exponentially
small wings), allowing for relatively quick data analysis. A problem
of those method, however, is that one can expect a degraded signal
to noise from the initial data, due to the use of compensated filters
that impose cancellation of part of the signal. This have not been
an important issue for two-point functions.

The $M_{ap}$ approach has been also applied to three point functions
\cite{2003astro.ph..7393J,2003astro.ph..2031P,2005A&A...431....9S}.
It has been showed that one can exhibit a summation procedure for
measured data that results into an estimation of the $M_{ap}$ filtered
bispectrum of the convergence field. Measurements on real data have
been performed. The quality of data being low and the degradation
of the signal-to-noise ratio inherent to $M_{ap}$ statistics results
in very large error-bars \cite{2003astro.ph..7393J,2003astro.ph..2031P}.

Owing to its simple relation with the convergence bispectrum and thus
to the matter distribution bispectrum, the $M_{ap}$ approach is certainly
a very appealing estimator of the weak lensing three-point function.
However, as it seems to require a high quality dataset, the need for
a simple and robust estimator of the weak-lensing three-point function
remains. The shear three-point functions $\Gamma_{i}$ can provide
such tool, but they still have a complicated dependency on the configuration.
Our goal in this section, is to use our analytical prescription in
order to propose a way to build an estimator with simpler properties,
yet avoiding as much as possible cancellations to preserve the signal-to-noise
ratio.

\subsection{Previous estimators}

Bernardeau, van Waerbeke and Mellier \cite{2003AA...397..405B} (hereafter
BvWM) proposed a simple estimator that exhibits some of the properties
discussed above. They studied the pseudo vector field $\boldfwm$\begin{equation}
\boldfwm(\ell_{1},\ell_{2},\ell_{3})=\left\langle \boldsymbol{\gamma}_{1}\left(\boldsymbol{\gamma}_{2}\cdot\boldsymbol{\gamma}_{3}\right)\right\rangle ,\label{eq:FWM:def}\end{equation}
 where all pseudo vectors $\boldsymbol{\gamma}$ are projected along
the direction given by $\boldsymbol{\ell}_{1}$. In such case, symmetry
does not imply that $\mathrm{\mathcal{F}}$ vanishes. BvWM were able
to evaluate $\mathcal{F}$ for special configurations of the three
points. They also computed it in the framework of the hierarchical
ansatz and measured it in N-body simulations. From their results,
it seemed that the parity-negative part of $\boldfwm$ was negligible
and that the parity-positive part was not changing sign in a small
region around $\theta_{2}\theta_{3}$. They empirically fitted the
shape of this region by an ellipse of focal points $\theta_{2}$,
$\theta_{3}$. In light of this, they proposed to use the averaging
on the ellipse of the parity positive component of $\boldfwm$ as
their estimator of the shear three point function.

The pseudo-vector $\mathcal{F}$ can be re-expressed as a combination
of the $\Gamma_{i}$. Let us place ourselves in the projection convention
defined by the direction of $\theta_{2}\theta_{3}$ (this choice of
projection is peculiar in the sense that it artificially breaks the
symmetry properties by distinguishing one of the points). With this
choice, the scalar product $\left(\boldsymbol{\gamma}_{2}\cdot\boldsymbol{\gamma}_{3}\right)$
reads\begin{equation}
\left(\boldsymbol{\gamma}_{2}\cdot\boldsymbol{\gamma}_{3}\right)=\gamma_{(2)+}\gamma_{(3)+}+\gamma_{(2)\times}\gamma_{(3)\times}.\end{equation}
 Multiplying the previous Eq. by $\gamma_{1}$ leads to the expression
of $\boldfwm$\begin{equation}
\boldfwm=\left(\begin{array}{l}
\gamma_{+++}+\gamma_{+\times\times}\\
\gamma_{\times++}+\gamma_{\times\times\times}\end{array}\right),\end{equation}
 which can be written in terms of the $\Gamma_{i}$\begin{equation}
\fwm=\frac{1}{2}\left(\Gamma_{2}+\Gamma_{3}\right).\label{eq:fwm:Gamma}\end{equation}
 Remember that the last equation holds only for the $\theta_{2}\theta_{3}$
projection convention; Eqs. (\ref{eq:Gamma0Rot}-\ref{eq:Gamma3Rot})
can be used to rotate the expression into some other projection convention.
\begin{figure}
\begin{center}\includegraphics[%
  bb=10bp 120bp 1320bp 1000bp,
  clip,
  width=1.0\columnwidth]{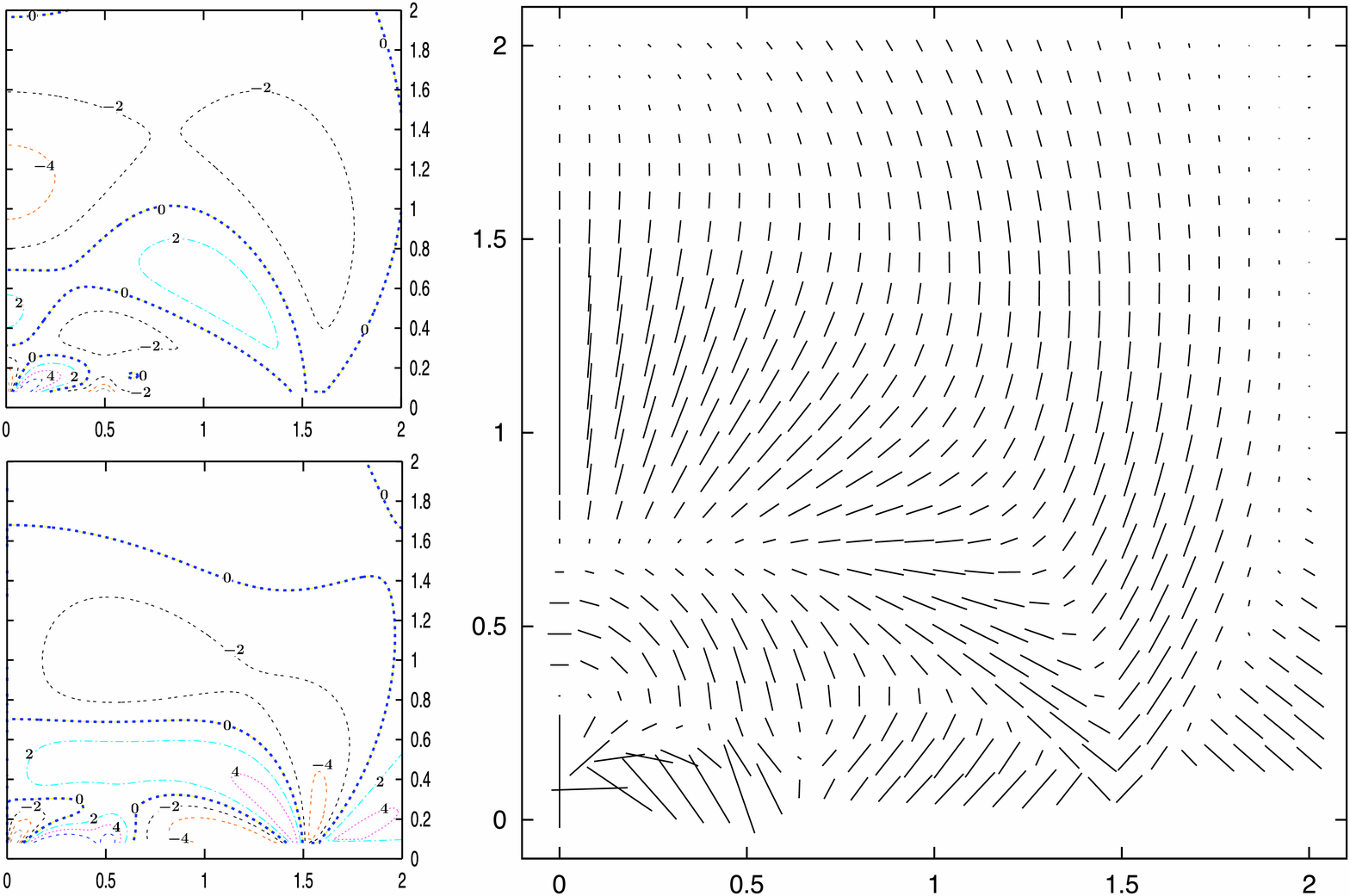}\end{center}

\caption{Predictions for the $\boldfwm$ estimator. Left panel presents the
real/parity positive (top) and imaginary/parity negative (bottom)
parts. Right panel shows the resulting pattern. The points $\theta_{2}$
and $\theta_{3}$ are fixed to $(0,-1/2)$ and $(0,1/2)$. The coordinate
gives the position of $\theta_{1}$. We only shows one fourth of the
plane. The rest of it can be determined from parity properties. $\fwm_{\times}$
goes go to zero on the x and y axis.\label{fig:FWMdensityplot}}
\end{figure}

Figure \ref{fig:FWMdensityplot} presents the $\boldfwm$ pseudo-vector
in the top right quadran of the plane, computed from our analytic
model, with index $n=2$. The base of triangle $(\theta_{2}\theta_{3})$
spans the range $[-1/2,1/2]$ on the x axis. Fig. \ref{fig:FWMdensityplot}
partly reproduce the result from \cite{2003AA...397..405B}. In particular,
even if the parity-negative part of the pseudo-vector $\fwm$ is smaller
than the parity positive component, it is not negligible. We also
explore the region where the parity positive component keeps the same
sign. Contour plots on the left panels of Fig. \ref{fig:FWMdensityplot}
shows that its shape is richer than that of an ellipse. Changing the
profile index $n$ modifies slightly the properties of $\boldfwm$.
The overall shape shown Fig. \ref{fig:FWMdensityplot} is conserved.
Modifications are concentrated around the $x$ axis, when the configuration
is nearly flat; in particular, the number of constant sign region
along $y_{\theta_{1}}=0$ changes as can be expected from Fig. \ref{fig:Gam:nvar}.
Finally, the amplitude of the parity-negative component decreases
slightly when the value of the profile index increases. This explains
the apparent discrepancy between BvWM results and ours. They are averaging
their measurement in numerical simulations on scales bigger that the
one probing the region where the halo profile is well described by
$n=2$. Moreover, the binning procedure can average out the parity
negative part.

Although ignoring the parity negative part and summing the parity
positive part over an ellipse is a good starting point, the BvWM estimator
can be improved by using the prediction of the expected shear pattern
as we now discuss.

\subsection{A new estimator: projecting measurements onto shear pattern templates}

An approach to avoid signal to noise cancellation due to positive
and negative contributions when combining different three-point functions
is to {}``project\char`\"{} the measurements of the $\Gamma_{i}$
directly onto the expected result (template) from analytic predictions
and build an estimator such as \begin{equation}
\mathcal{D}_{0}\equiv\sum_{i}\tilde{\Gamma_{i}}\Gamma_{i}^{*},\end{equation}
 $\tilde{\Gamma_{i}}$ being the analytical predictions, the star
denoting the complex conjugate. Note that $\mathcal{D}_{0}$ now depends
on our choice of profile index $n$. One can do even better, assuming
that the covariance of the $\Gamma_{i}$ is known, by building a {}``minimum
variance'' estimator of the three-point function \begin{equation}
\mathcal{D}_{1}\equiv\sum_{i,j}\tilde{\Gamma_{i}}\left\langle \Gamma_{i}\Gamma_{j}^{*}\right\rangle ^{-1}\Gamma_{j}^{*},\end{equation}
 where $\left\langle \Gamma_{i}\Gamma_{j}^{*}\right\rangle ^{-1}$
denotes the inverse of the covariance matrix. $\mathcal{D}_{1}$ is
only a minimum variance estimator to the extent that its distribution
can be approximated to be Gaussian.

It has been shown that to a good approximation, the covariance of
the three-point function of the weak lensing shear can be evaluated
by restricting the computation to the Gaussian contribution \cite{2003astro.ph..4034T}.
However, even if we restrict ourselves to the Gaussian contribution,
the covariance matrix is difficult to compute analytically. Indeed,
as discussed above the geometrical properties of the shear complicate
the calculation of the three-point functions, and the situation is
even worse here, as we have to integrate over all possible orientations
and positions of two identical triangles. To avoid such complication,
we evaluate the covariance matrix by measuring it in our numerical
simulations (using 40 realizations, see Appendix B). Even in this
case the resulting evaluation of the covariance matrix is quite noisy,
which somewhat reduces the reliability of $\mathcal{D}_{1}$, but
nevertheless provides an estimation of the expected signal-to-noise
improvement that can be expected from a better determination of the
covariance of the $\Gamma_{i}$.

\begin{figure}
\begin{center}\includegraphics[%
  bb=10bp 65bp 770bp 555bp,
  clip,
  width=1.0\columnwidth]{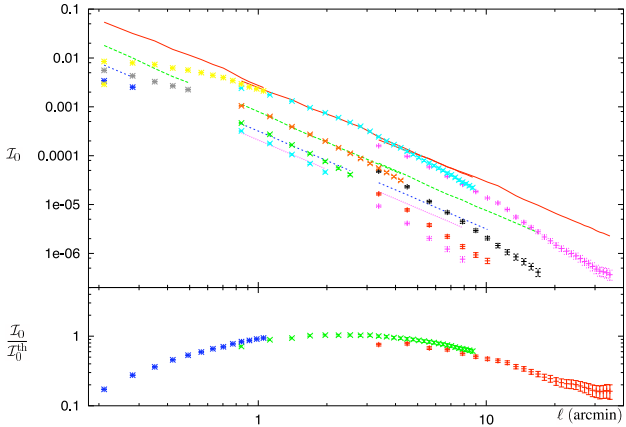}\end{center}

\caption{$\mathcal{I}_{0}$ measured in our three datasets.\label{fig:I0}
Results for all datasets are combined into this composite figure.
In the top panel, top to bottom set of points corresponds to measurement
for $\ell_{2}=r\ell_{3}$ for $r=1,2,3,4$. The corresponding theoretical
prediction for $n=2$ are shown on top of the points for $r=1$ (solid
line), $r=2$ (long dashed), $r=3$ (dashed) and $r=4$ (dotted).
Bottom panel shows the ratio between the measurement and theoretical
prediction for $r=1$.}
\end{figure}

\begin{figure}
\begin{center}\includegraphics[%
  bb=10bp 65bp 770bp 555bp,
  clip,
  width=1.0\columnwidth]{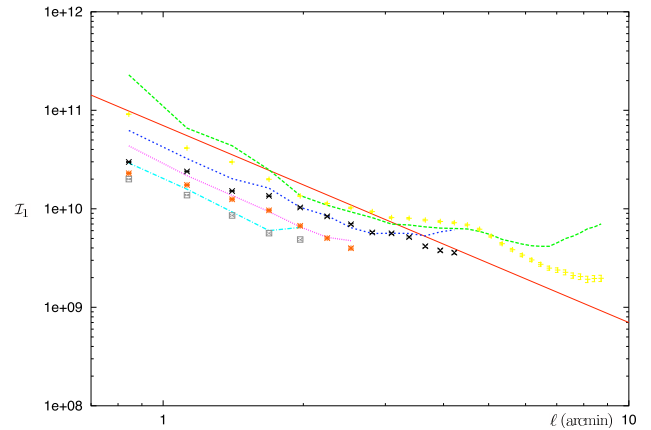}\end{center}

\caption{$\mathcal{I}_{1}$ measured in our medium resolution dataset. The
solid line is a power law proportional to $1/\ell^{2}$ and shows
the expected scaling. Other curves are semi analytical results (the
covariance matrix has been evaluated from the simulations). They correspond
to $r=1$ (long dashed), $r=2$ (dashed), 3 (dotted), $r=4$ (dot-dashed).
Measurements partially agree with the semi-analytical predictions
\label{fig:I1} }
\end{figure}

The $\mathcal{D}_{0}$ and $\mathcal{D}_{1}$ estimators depend on
the configuration of the three points. We reduce this dependence by
summing the estimators over a set of configurations. This sum is similar
to the approach taken by BvWM, where they integrated over all observed
configurations in a small ellipse. In our case there is no particular
choice of summation region, as we are guaranteed to avoid cancellations
as long as the effective profile index is close to the one of our
template ($n=2$, our fiducial choice). To simplify the process of
summation, we take advantage of the fact that our measurements are
already binned by length and opening angle, thus we sum configurations
along the opening angle of the triangle, $\psi_{1}$, for a given
ratio of lengths $\ell_{2}/\ell_{3}$. The overall length of the triangle
should be absorbed into the scaling relation (\ref{eq:scaling}) when
it holds. Thus we define \begin{equation}
\mathcal{I}_{0}(\ell,\ell_{2}/\ell_{3})=\int\dd\psi_{1} \mathcal{D}_{0}(\psi_{1},\ell,\ell\times\ell_{2}/\ell_{3}),\end{equation}
 and the corresponding $\mathcal{I}_{1}$, for the estimator $\mathcal{D}_{1}$.
When the effective profile index is close to $n=2$, Eq. (\ref{eq:scaling})
implies that $\mathcal{I}_{0}(\ell,\ell_{2}/\ell_{3})$ obeys the
following relation\begin{equation}
\mathcal{I}_{0}(\lambda\ell,\ell_{2}/\ell_{3})=\frac{1}{\lambda^{2}}\mathcal{I}_{0}(\ell,\ell_{2}/\ell_{3}).\label{eq:scal:I0}\end{equation}
 In other words, we expect the estimators to behave like power laws.
Figure \ref{fig:I0} and \ref{fig:I1} present the measurements of
$\mathcal{I}_{0}$ and $\mathcal{I}_{1}$ with the template profile
index $n=2$. Error bars correspond to the variance among different
realizations. The integral over the opening angle has been cut-off
at $\psi_{1}<\pi/9$ and $\psi_{1}>8\pi/9$ to avoid errors arising
from binning. The measurements are compared with analytic predictions
obtained by mimicking the measurement procedure, i.e. the integrals
of $\mathcal{D}_{0}^{\textrm{th}}=\sum_{i}\tilde{\Gamma_{i}}\tilde{\Gamma_{i}}^{*}$(resp.
$\mathcal{D}_{1}^{\textrm{th}}=\sum_{i,j}\tilde{\Gamma_{i}}\left\langle \Gamma_{i}\Gamma_{j}^{*}\right\rangle ^{-1}\tilde{\Gamma_{j}^{*}}$)
are computed with the same cutoff on $\psi_{1}$ and using the same
sampling in $\psi_{1}$ resulting from the binning of our dataset.
This explains the slight departure from a power-law in the theoretical
curves in Fig. \ref{fig:I0}. The normalization free parameter in
the theoretical model has been set so as to maximize the agreement
between the measured $\Gamma_{2}$ and its analytic estimation for
isoceles configuration at $1$ arc-min. The large difference in normalizations
between $\mathcal{I}_{0}$ and $\mathcal{I}_{1}$ is due to the fact
that we did not normalize the covariance matrix; it accounts for the
inverse of the covariance matrix determinant. The estimators $\mathcal{I}_{0}$
and $\mathcal{I}_{1}$ have been measured for isoceles triangles,
as well as for some elongated configurations $\ell_{2}/\ell_{3}=2,3$
and $4$. Due to our data analysis strategy only a few elongated configurations
are available (see discussion in Appendix \ref{sec:Nbody}). Figure
\ref{fig:I0} shows the results for all our dataset, whereas Figure
\ref{fig:I1} focuses only on the medium scale dataset, where the
effective profile index is expected to be close to $n=2$.

As expected, $\mathcal{I}_{0}$ is very close to its theoretical estimation
between $1'$ and $4'$ where the effective profile index is close
to $n=2$. The agreement is very good for isoceles triangles as well
as for the elongated configurations. The agreement between the theoretical
and measured $\mathcal{I}_{1}$ is not as good. Note that Figure \ref{fig:I1}
seems nevertheless to indicate a behavior similar to a power law,
corresponding to $n_{\textrm{eff}}=2$. Clearly, here we are sensitive
to the noise in our estimation of the covariance matrix. Even with
our cut in $\psi_{1}$, $\mathcal{D}_{1}$ still gets a significant
contribution from flattened configurations, where the covariance matrix
terms are large. The discreetness errors induced by the data binning
are also responsible for part of the discrepancy. Measurements for
elongated configurations ($\ell_{2}/\ell_{3}>1$), where discreetness
effects are less important by construction, show a better agreement
with the analytic estimations. For scales out of the $1-4$ arc-min
ranges, the theoretical pattern is no longer valid, and the projections
$\mathcal{D}_{0}$ and $\mathcal{D}_{1}$ decrease. We can see this
behavior in Figure \ref{fig:I0}.

\begin{figure}
\begin{center}\includegraphics[%
  bb=10bp 65bp 770bp 555bp,
  clip,
  width=1.0\columnwidth]{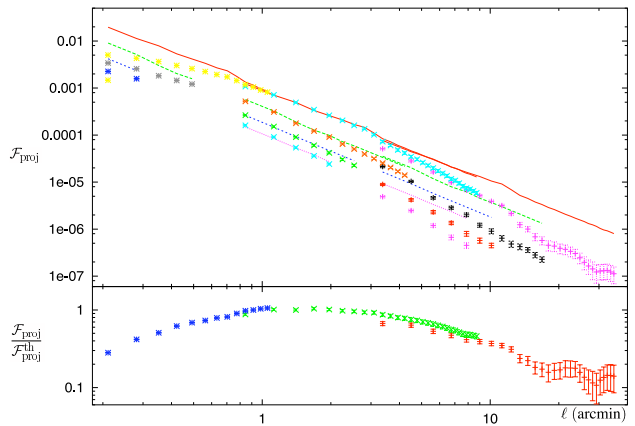}\end{center}

\caption{Same as Fig. \ref{fig:I0} for our improved version of the BvWM estimator.\label{fig:Fproj} }
\end{figure}

We now estimate the improvement of signal to noise in our estimator
compared to the one proposed by BvWM. More precisely, we measure the
pseudo-vector $\fwm$, as a function of the configuration, project
it on our analytical model and sum it along the opening angle of the
triangle, in a similar way to what we have done for $\mathcal{I}_{0}$
and $\mathcal{I}_{1}$. Note that this is an enhancement of BvWM method
since we are using both components of the pseudo vector $\fwm$. Indeed,
since we are projecting on analytical predictions, the parity-negative
part will no longer average to zero. Figure \ref{fig:Fproj} presents
the measurements in the simulation as well as the theoretical prediction
and can be directly compared with Figure \ref{fig:I0}.

\begin{figure}
\begin{center}\includegraphics[%
  bb=10bp 65bp 770bp 555bp,
  clip,
  width=1.0\columnwidth]{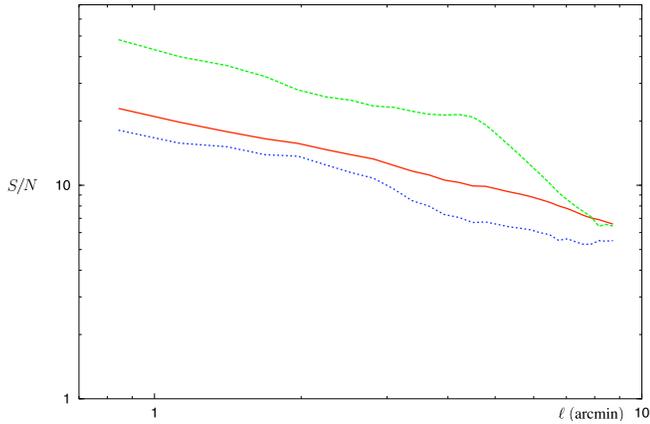}\end{center}

\caption{Signal to noise for $\mathcal{I}_{0}$ (solid line), $\mathcal{I}_{1}$
(dashed) and our improved version of the BvWM estimator (dotted).
Only isoceles configurations have been taken into account. Data are
projected on the $n=2$ model predictions. Only the medium resolution
datasets are used.\label{fig:S2N} }
\end{figure}

Finally, Figure \ref{fig:S2N} shows the signal to noise ratio for
the different estimators, evaluated in our simulations. We do not
take into account experimental noise such as the distribution of intrinsic
ellipticity of the galaxies or the uneven distribution of sources.
Our estimation is thus dominated by the cosmic variance. The shot
noise term due to the intrinsic orientation of the galaxies will mainly
contribute as a non correlated source of noise at small scales. For
the simplified case of the convergence three-point function of equilateral
triangles, it has been shown that this term contributes at scales
smaller than 1 arc-minute \cite{2003astro.ph..4034T}. As expected,
$\mathcal{I}_{1}$ has the best S/N ratio, nearly twice better than
$\mathcal{I}_{0}$. This however degrades quickly as the effective
profile index leaves the $n_{\textrm{eff}}\sim2$ region. The S/N
ratio of $\mathcal{I}_{0}$ is about 30\% better than for our improved
BvWM estimator. This is not unexpected, Eqs. (\ref{eq:fwm:Gamma})
shows that $\fwm$ can be computed with $ $only two of the $\Gamma_{i}$,
whereas $\mathcal{I}_{0}$ uses all four of them. If $\Gamma_{i}$
are uncorrelated, this should correspond to a $\sqrt{2}$ degradation
of the signal-to noise ratio, which is close to the 30\% we measure
here.%

Recall that our improved BvWM estimator uses the full $\fwm$ pseudo-vector.
Using only the parity positive coordinate will further reduce the
signal to noise by another factor $\sqrt{2}$ if its two components
are uncorrelated.

\section{Conclusion and Discussion}

We have described a simplified analytic model that can be used to
compute the three-point functions of the shear. This model is inspired
by halo models, and only considers the one-halo contribution of a
spherical potential of power law profile. The free parameters of our
models are the profile index $n$ and a normalization. We have used
this model to investigate some geometrical properties of the shear
three-point function. In particular, we have shown that there is no
preferred projection choice that will reduce the number of independent
three-point functions.

We have compared the predictions of this model with results from N-body
simulations. The predicted and measured three-point functions of the
shear where shown to be in good agreement. In particular, we have
shown that the approximation made on the profile index is sufficient
to predict the shear three points function in a reasonable range of
scales. The isothermal profile case, $n=2$, corresponding to scales
from $1'$ to $4'$. These scales correspond indeed to the scales
where the dominant halos at redshift $z=0.4$ are seen with a local
profile index $n=2$. As expected, the smaller scales exhibited a
behavior compatible with an index $n<2$, while larger scales where
compatible with $n>2$. Our model allows for computations with a modified
profile index, and it can thus be used at smaller or larger scales.
However, at scales bigger than $4'$, the one-halo dominant contribution
model will break down, and one should take into account two and three
halo terms \cite{2003MNRAS.340..580T}.

Using the agreement between our model and synthetic data, we proposed
an optimized measurement of the cosmic shear three-point function.
Our method trades the easier cosmological analysis allowed by $M_{ap}$
reconstructions methods for a better signal to noise of the measurement
by avoiding cancellations. We use the model predictions to compute
{}``optimal'' weighted sums of the eight three-point functions.
Contrarily to the $M_{ap}$ statistic, this kind of estimator is local
and is not affected by the shape of the survey. These methods can
be seen as a refined version of the one proposed by BvWM \cite{2003AA...397..405B}.

We computed two such estimators, the first a simple projection on
our theoretical model, the other taking into account an estimation
of the covariance matrix of the shear three-point functions. We compared
them with an improved version of the estimator implemented by BvWM.
Our estimators perform much better than the improved BvWM. The minimal
variance estimator lets us expect more than a factor of two gain in
the signal to noise in the best case.

With future space-based experiments, the quality of cosmic shear data
will greatly increase and the loss of signal to noise inherent to
compensated filters will be a not too high a price to pay for accessing
to the filtered three-point function of the convergence. In the meantime,
we believe that methods using projections of data onto theoretical
predictions, as the one we proposed here, will be an interesting alternative.
Improvement to what we proposed here will be doubtless needed. In
our analysis, we first measured the three-point functions and then
projected them on analytic templates, increasing the errors due to
discreteness of the binning. We saw that this error has a significant
impact on elongated isoceles configurations. Projecting the data as
they are measured can solve this problem. Using a simple projection
already improves the situation compared to what has been done before.
The minimal variance estimator promises an even better ability to
detect the shear three-point functions. Results from this estimator
are yet difficult to forecast as our evaluation of the covariance
matrix is somewhat noisy. More numerical simulations will be needed
to quantify this better.

Improving the choice of effective halo profile index will also result
in a net improvement, since by restricting the index to that of an
isothermal sphere, $n=2$, we were only able to improve the efficiency
of the three point function estimator around 1-4 arcminutes. This
is not a restriction of the model, since it can be easily extended
to any other index profile. More generally, given any model for the
three-point functions one can compute the {}``optimal\char`\"{} estimators
we define here.

Finally, we have not investigated how our method can be used to obtain
information on the underlying cosmological model. With our simple
model, all the cosmological information is encoded in the free normalization
parameter. Further work, comparing our simple model, with a full halo
model will be needed to investigate this point.

\acknowledgements  The numerical simulations and analysis presented
here have been run at the NYU Beowulf cluster supported by NSF grant
PHY-0116590. The authors wish to thank M. Zaldarriaga, F. Bernardeau and Y. Mellier for useful discussions and comments.
\appendix

\section{Projection onto the opposite side\label{sec:app:projections}}

Here we describe the computations of the $\trig_{\mu}(2\phi_{i})$
factor for the opposite side projection described in section 3.1.
Remember that the angles $\phi_{i}$ are defined by\[
\cos(\phi_{i})\equiv\frac{\left(\boldsymbol{u}-\boldsymbol{\theta}_{i}\right)\cdot\boldsymbol{\ell}_{i}}{\left|\boldsymbol{u}-\boldsymbol{\theta}_{i}\right|\ell_{i}}.\]
 The computation of $\trig_{+}(2\phi_{i})=\cos(2\phi_{i})$ is the
simplest. Defining $\boldsymbol{d}_{i}\equiv\boldsymbol{\theta_{i}}-\boldsymbol{u}$,
one gets\[
\cos(\phi_{i})=\frac{\boldsymbol{\theta_{i}}\cdot\boldsymbol{\ell_{i}}-\frac{1}{2}\left(\theta_{i+1}^{2}-d_{i+1}^{2}-\theta_{i+2}^{2}+d_{i+2}^{2}\right)}{d_{i}\ell_{i}},\]
 which, using the identity \[
2\boldsymbol{\theta_{i}}\cdot\boldsymbol{\ell_{i}}-\theta_{i+1}^{2}+\theta_{i+2}^{2}=\ell_{i+1}^{2}-\ell_{i+2}^{2}\]
 reduces to\[
\cos(\phi_{i})=\frac{d_{i+1}^{2}-d_{i+2}^{2}+\ell_{i+1}^{2}-\ell_{i+2}^{2}}{2d_{i}\ell_{i}}.\]
 The last step needed to obtain $\cos(2\phi_{i})$ is straightforward,
using $\cos(2\phi_{i})=2\cos(\phi_{i})^{2}-1$, one gets the final
expression of $\cos(2\phi_{i})$ as a polynomial of the 4th degree
in $d_{i+1}$ and $d_{i+2}$ and second degree in $d_{i}$, divided
by $d_{i}^{2}\ell_{i}^{2}$\begin{eqnarray*}
\trig_{+}(2\phi_{1}) & = & \frac{1}{2d_{i}^{2}\ell_{i}^{2}}\left[d_{i+1}^{4}+d_{i+2}^{4}-2d_{i+1}^{2}d_{i+2}^{2}-2d_{i}^{2}\ell_{i}^{2}\right.\\
 &  & +2\left(d_{i+1}^{2}+d_{i+2}^{2}\right)\left(\ell_{i+1}^{2}-\ell_{i+2}^{2}\right)\\
 &  & \left.+\left(\ell_{i+1}^{2}-\ell_{i+2}^{2}\right)^{2}\right].\end{eqnarray*}

The parity negative terms require a little more work. One can use
the identity $\sin(2\phi_{i})=2\cos(\phi_{i})\sin(\phi_{i})$. The
difficulties lie in the computation of $\sin(\phi_{i})$. One can
use the fact that\begin{equation}
\sin(\phi_{i})=\frac{\left(\boldsymbol{u}-\boldsymbol{\theta}_{i}\right)\wedge\boldsymbol{\ell}_{i}}{\left|\boldsymbol{u}-\boldsymbol{\theta}_{i}\right|\ell_{i}}\cdot\boldsymbol{z},\label{eq:sinphii1}\end{equation}
 and re-express the unit vector $\boldsymbol{z}$ as \begin{equation}
\boldsymbol{z}=\frac{\boldsymbol{\ell}_{i+2}\wedge\boldsymbol{\ell}_{i}}{\ell_{i+2}\ell_{i}\sin(\psi_{i+1})}.\end{equation}
 Finally, a careful invocation of the identity\begin{eqnarray}
\left(\boldsymbol{A}\wedge\boldsymbol{B}\right)\cdot\left(\boldsymbol{C}\wedge\boldsymbol{D}\right)+\left(\boldsymbol{B}\cdot\boldsymbol{B}\right)\left(\boldsymbol{C}\cdot\boldsymbol{D}\right) & =\\
 &  & \hspace{-5cm}\left(\boldsymbol{A}\wedge\boldsymbol{C}\right)\cdot\left(\boldsymbol{B}\wedge\boldsymbol{D}\right)+\left(\boldsymbol{A}\cdot\boldsymbol{C}\right)\left(\boldsymbol{B}\cdot\boldsymbol{D}\right)\nonumber \end{eqnarray}
 allows us to re-express Eq. (\ref{eq:sinphii1}) as \begin{eqnarray}
\sin(\phi_{i}) & = & \frac{1}{d_{i}\ell_{i}^{2}\ell_{i+2}\sin(\psi_{i+1})}\left\{ -\left(\boldsymbol{u}-\boldsymbol{\theta}_{i}\right)\cdot\boldsymbol{\ell}_{i}\,\boldsymbol{\ell}_{i}\cdot\boldsymbol{\ell}_{i+2}\right.\nonumber \\
 &  & \hspace{.3cm}\left.+\left[\left(\boldsymbol{u}-\boldsymbol{\theta}_{i+2}\right)+\left(\boldsymbol{\theta}_{i+2}-\boldsymbol{\theta}_{i}\right)\right]\cdot\boldsymbol{\ell}_{i+2}\,\ell_{i}^{2}\right\} \end{eqnarray}
 which can be written as a function of $\cos(\phi_{i+2})$ and $\cos(\phi_{i})$
and a term that only depends on the configuration\begin{eqnarray}
\sin(\phi_{i}) & = & \frac{1}{d_{i}\ell_{i}^{2}\ell_{i+2}\sin(\psi_{i+1})}\left(\cos(\phi_{i+2})d_{i+2}\ell_{i+2}\,\ell_{i}^{2}\right.\nonumber \\
 &  & \hspace{.3cm}\left.+\cos(\phi_{i})d_{i}\ell_{i}\,\boldsymbol{\ell}_{i}\cdot\boldsymbol{\ell}_{i+2}+\boldsymbol{\ell}_{i+1}\cdot\boldsymbol{\ell}_{i+2}\ell_{i}^{2}\right).\end{eqnarray}
 This last expression can be expressed only in terms of $\ell_{j=1,2,3}$
and $d_{j=1,2,3}$ as the ratio between a 4th degree polynomial in
the $d_{j=1,2,3}$ divided by $d_{i}$, the coefficient being functions
of the $\ell_{j=1,2,3}$. After simplifications, the final result
for $\trig_{\times}(2\phi_{i})=\sin(2\phi_{i})$reads\begin{eqnarray}
\trig_{\times}(2\phi_{i}) & = & \frac{\left(d_{i+1}^{2}-d_{i+2}^{2}+\ell_{i+1}^{2}-\ell_{i+2}^{2}\right)}{2d_{i}^{2}\ell_{i}^{2}\sqrt{4\ell_{i+1}^{2}\ell_{i+2}^{2}-\left(\ell_{i}^{2}-\ell_{i+1}^{2}-\ell_{i+2}^{2}\right)^{2}}}\nonumber \\
 &  & \times\left[-2d_{i}^{2}\ell_{i}^{2}+d_{i+1}^{2}\left(\ell_{i}^{2}+\ell_{i+1}^{2}-\ell_{i+2}^{2}\right)\right.\nonumber \\
 &  & +d_{i+2}^{2}\left(\ell_{i}^{2}-\ell_{i+1}^{2}+\ell_{i+2}^{2}\right)+\ell_{i+1}^{4}+\ell_{i+2}^{4}\nonumber \\
 &  & \left.-\ell_{i}^{2}\left(\ell_{i+1}^{2}+\ell_{i+2}^{2}\right)+2\ell_{i+1}^{2}\ell_{i+2}^{2}\right]\end{eqnarray}

\section{Numerical simulations\label{sec:Nbody}}

We used the GADGET \cite{2001NewA....6...79S} code to evolve 24 realizations
of the large scale structure in a small section of a FRLW universe
with parameters $\Omega_{m}=0.3$, $\Omega_{\Lambda}=0.7$ $h=0.7$.
The initial conditions were imposed at redshift $z=50$ using second-order
Lagrangian perturbation theory~\cite{1998MNRAS.299.1097S}. The initial
power spectrum used was obtained using the B\&E \cite{1984ApJ...285L..45B}
fitting formula and normalized to $\sigma_{8}=0.9$ at $z=0$ in linear
theory. The boxes we considered are cubes of $(100\,\textrm{Mpc}/h)^{3}$
containing $200^{3}$ particles. This choice has been made has a trade-off
between speed and accuracy. The simulation were run on $24$ nodes
of our cluster. For each realization, we took a snapshot of the large
scale structures every $100$~Mpc/$h$.

\subsection{Building the light-cone}

We use these snapshots to compute the weak lensing effect at a redshift
of unity on a square light cone. At $z=1$ the sides of the boxes
represent $2.47$ degrees. We will only produce $2.4^{2}$ square-degrees
patches of the sky.

The line of sight to the sources, is built by tilling $17$ snapshots
at different redshifts. We use this construction to create more pseudo
realizations of the lensing effect that we have different realizations
of the density field. We follow a very similar method to the one exposed
by White and Hu in \cite{2000ApJ...537....1W}.

While building each line of sight, we will make sure that each N-body
realization is only used once. To increase the randomness of our lensing
pseudo-realization, we translate each snapshot by a random vector,
taking advantage of the periodic boundary condition of the boxes.
Moreover, we trace the path of the photon along a random direction
in the box, and not along the axis direction. Note however that all
rotation angles are not allowed if one wants to avoid tracing twice
the same structures in a given box. We take this problem into account
when picking the ray-tracing direction.

The projected mass density is then built as follows. For each of our
$17$ snapshots, after translation and rotation, we build a list of
particles belonging to the light cone. This list of particles is flattened
onto a 2D density map, using a Cloud-in-Cell algorithm. This 2D map
is then multiplied by the efficiency window function of the lensing
effect to provide a map of the convergence $\kappa$. These $\kappa$
slices are added to produce the final lens effect on the source plane.
The shear field $\boldsymbol{\gamma}$ on each of these synthetic
$\kappa$ maps is obtained by numerically solving Eq. (\ref{eq:gammadef})
with a FFT~\cite{2001MNRAS.322..918V}.

Note that in our computation of the lens effect we neglected some
secondary effects. For example our ray-tracing scheme is strictly
restricted to the Born approximation. We thus assume here that the
lens effect computed along the unperturbed path of the photon gives
a good evaluation of the effect. This as been tested in numerous work
before \cite{2001MNRAS.322..918V,2000ApJ...530..547J}. Doing this,
we neglect the lens lens-coupling which is known to produce non-zero
corrections to the three-point function of the convergence field.
This correction is expected to be small \cite{2001MNRAS.322..918V},
and we will neglect it here as our goal is mainly to evaluate the
validity of our simplified halo model.

The number of slices of our light-cone is more of a concern. A naive
evaluation of the impact of this choice can be made by comparing a
step summation of the filtered lensing power spectrum with its full
integration. For an Einstein-deSitter universe, assuming a power law
matter density fluctuation, we thus have to compare step summation
and integration of the function\begin{equation}
R(t)=x^{4-n}(x-1)^{2}.\end{equation}
 For the spectral index $n\sim-\frac{3}{2}$, the difference between
the summation and integration falls below one percent as soon as the
number of slices is bigger than 5. Vale and White \cite{2003astro.ph..3555V}
recently performed a less naive evaluation of the same effect using
N-body simulations. They showed that with simulations comparable to
ours ($300$Mpc and $512^{3}$particles) one can expect about 5\%
discrepancy on the power spectrum between a computation with a slice
every $\Delta s=125$Mpc (we are slicing every $100$Mpc ) and one
with $\Delta s=25$Mpc.%
\begin{figure}
\begin{center}\includegraphics[%
  bb=40bp 60bp 750bp 565bp,
  clip,
  width=1.0\columnwidth]{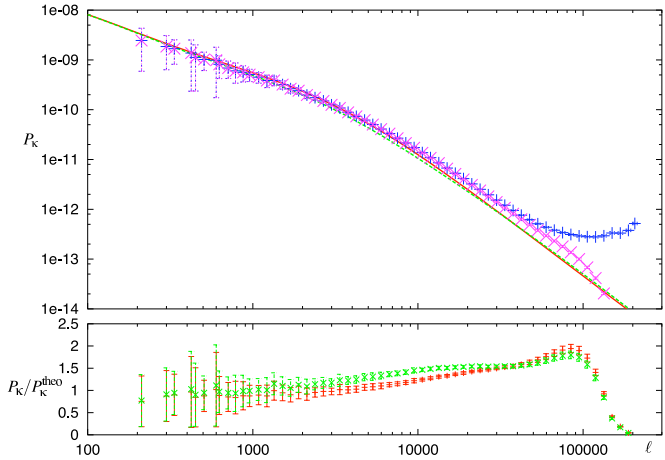}\end{center}

\caption{Power spectrum of $\kappa$ measured on the 40 lensing pseudo-realization.
Lines are the predicted power spectrum using the PD \cite{1996MNRAS.280L..19P}
(dashed) or halofit \cite{2003MNRAS.341.1311S} (solid) prescription
for the small scales. Points are the measurements in the simulations,
without smoothing (crosses), and with smoothing (xs) to remove the
Poisson noise (see text). Bottom panel shows the ratio between the
smoothed measured power spectrum and the two predictions using PD
(xs) and halofit (crosses). \label{cap:Pkappamesured} }
\end{figure}

\subsection{Resolution }

During the Cloud-in-Cell remapping of the particle list, the interpolation
was done on a $2048^{2}$ grid. The maximum resolution of our simulation
is thus $4.21$ arc-sec. However, at this scale we expect to probe
the region where shot noise starts to dominate {[}see equation (24)
from \cite{2000ApJ...530..547J}{]}. To reduce shot noise contributions
we smooth the $\kappa$ maps by a 2 pixel wide Gaussian window.

The average over our 40 realizations of the $\kappa$ power spectrum
is presented Fig. \ref{cap:Pkappamesured} and compared with the theoretical
one obtained by two ansatz of the non-linear density power spectrum.
The power spectrum is somewhat higher than the analytical predictions
in the non-linear regime. A similar behavior can be seen in the 3D
power spectrum of our simulations.

\subsection{Measurement strategy}

The measurement of two and three point functions in real space requires
significant computing time. These operations scale respectively as
$N^{2}$ and $N^{3}$ where $N=2048^{2}$ in our case. To reduce the
time needed to perform the computation, we will only probe configurations
up to a cut-off scale. This reduces the amount of computer time needed
to perform the measurement, but prevents us to access to a broad range
of scales.

To reduce the computation cost yet preserving the ability to measure
the three-point functions over a large range of scales, we measure
the two- and three-point functions at large scales on degraded resolution
maps obtained by top-hat filtering and regridding the synthetic shear
fields. For each field, we produced three datasets; one with the nominal
resolution of $2048^{2}$ pixels representing of $4.22^{2}$ arc-sec$^{2}$,
one with a four time degraded resolution ($512^{2}$ pixels, $16.9^{2}$
arc-sec$^{2}$) and the last with a sixteen time degraded resolution
($128^{2}$ pixels, $1.125^{2}$ arc-min$^{2}$). For the two-point
functions, the computation cost is lower and we are able, for the
two last resolution sets to measure it without cut-off scale; we apply
a cut-off only for the biggest map and only consider scales smaller
than a 32th of the map size.

For the three-point functions we only explore a small region of the
three points configuration space. Table \ref{tab:whatever} presents
the scales probed by each of our datasets.%
\begin{table}
\begin{center}\begin{tabular}{c|c|c|c}
&
 max&
 medium&
 min\tabularnewline
\hline
$\Delta\ell$&
 $4.22''$&
 $16.9''$&
 $1.125'$\tabularnewline
\hline
$\ell_{\textrm{max}}$&
 $67.5''$&
 $9.0'$&
 $36'$\\
\tabularnewline
\end{tabular}\end{center}

\caption{Description of the three datasets. \emph{Max} is the best resolution
dataset, \emph{min} the worst one. Thorough this article, we will
mainly use the medium resolution.\label{tab:whatever}}
\end{table}

The cut-off scales have been chosen so as to have each measurement
to require about ten days of computation time of one node of the our
cluster.

We take into account the cut-off scale we added to optimize the measurement.
Since the data are gridded, we can precompute for any given point
all the positions of the couples of points which will create a valid
configuration. By valid configuration we mean any triangle that fits
the requirement of the cut-off scale, whose three point function measured
with a given projection convention has a meaningful result (i.e. for
the center of mass projection, we throw out configurations where the
center of mass is one of the vertices of the triangle) and is such
that a given configuration is only seen once when varying the initial
position. This last requirement can be fulfilled by requiring that
for any couple of points \begin{eqnarray}
x_{1}\geq x_{i},\, x_{2}\geq x_{1}\, y_{1}>y_{i} & \textrm{if} & x_{1}=x_{i},\nonumber \\
y>y_{1} & \textrm{ if } & x_{2}=x_{1},\end{eqnarray}
 provided that the initial point will go through the grid increasing
its $y_{i}$ position first.

We build a table containing for each of such configuration, the offset
of the points positions, as well as the projector vectors on the $+$
and $\times$ direction for each point of the triangle. Populating
this table is an expensive task as it goes as $l^{4}$ in time and
memory, where $l$ is our cut-off scale. Once this initialization
done, however, we just have to traverse the shear map and apply for
each point the rules contained in the initialization table.

\subsection{Shear two-point functions}

We measured the two point functions of the shear $\xi_{\pm}$\begin{equation}
\xi_{\pm}(\theta)=\left\langle \gamma_{+}\gamma_{+}(\theta)\right\rangle \pm\left\langle \gamma_{\times}\gamma_{\times}(\theta)\right\rangle ,\label{eq:xidef}\end{equation}
 where the shear pseudo-vector is projected to the $+$ and $\times$
directions defined by the vector linking the two points and the same
vector rotated by $\pi/4$ (see section \ref{sub:shear2}). The relation
between the shear two-point functions and the power spectrum of the
convergence is well known \cite{1992ApJ...388..272K}\begin{equation}
\xi_{\pm}(\theta)=\int\dd l\, l\, P_{\kappa}(l)\, J_{0,4}(l\theta).\label{eq:xifromPkappa}\end{equation}
\begin{figure}
\begin{center}\includegraphics[%
  bb=70bp 60bp 760bp 550bp,
  clip,
  width=1.0\columnwidth,
  keepaspectratio]{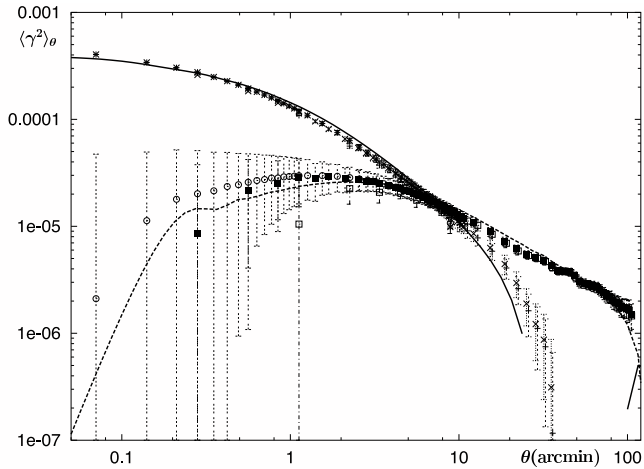}\end{center}

\caption{The shear two-point functions in the simulations. Lines are the predicted
power spectrum, taking into account the survey size and resolution.
Solid line is $\xi_{+}$, dashed, $\xi_{-}$. Points show the measurements
in the 40 realizations, for our three different datasets. The open
circle (resp. plain squares, open squares) shows $\xi_{-}$ and regular
pluses (resp crosses,dashed pluses) $\xi_{+}$, measured in the max
(resp medium, min) resolution datasets. \label{cap:xi}}
\end{figure}

Figure \ref{cap:xi} presents the measurement of $\xi_{\pm}$ in our
simulations as well as analytical predictions. It can be surprising
to see that at a scale corresponding to only a fourth of the box scale,
the two point shear correlations, and especially the $\xi_{+}$ one,
differ greatly from the usual analytical prediction. This corresponds
to the effect of the finite survey size. Even at a scale four times
smaller than our survey size, we are already suffering from the lack
of large-scale correlations. We can reproduce it in our analytical
prediction as follows: we are filtering the convergence power spectrum
by a Bessel $J_{0}$ or $J_{4}$ function. $J_{0}$ and $J_{4}$ are
maximum when their argument is small. This means that when the angular
scale is large, the filter overweights the small $l$ in $P(l)$.
This is more severe for $J_{0}$ as its envelope decrease more quickly
than the one of $J_{4}$. Even though $l\, P(l)$ quickly decrease
when $l$ goes to $0$, it amounts to a difference that can be observed
in the comparison between analytical computations and measurements.
If we artificially cut the analytical power spectrum in our computation
to reproduce the absence of scale larger than $\sqrt{2}\times2.4$
arc-minutes in our simulation, we obtain a far better agreement with
our measurement.

Similarly, the small scale behavior of $\xi_{\pm}$ is modified by
the filtering we applied to our synthetic field by downgrading their
resolution. The analytic predictions, once this filtering is included,
are in good agreement with our numerical results.

We are thus quite confident than our implementation of the measurement
is sound and that the resolution degradation procedure gives meaningful
results.

\bibliographystyle{aa}
\bibliography{skew}

\end{document}